\documentclass[11pt]{article}

\usepackage{ulem} 
\usepackage{hyperref}
\hypersetup{
    colorlinks=true,
    linkcolor=blue,
    filecolor=magenta,      
    urlcolor=blue,
}

  \usepackage{geometry}
\usepackage{amsmath,amssymb,amsthm}
\usepackage{graphics,epsfig,calc}
\usepackage{amsmath}

\usepackage{latexsym,epsfig,bm,amssymb}
\usepackage{xcolor}
\usepackage{amsthm,mathrsfs}

\usepackage{mathptmx}

\DeclareSymbolFont{AMSb}{U}{msb}m{n}
\DeclareSymbolFontAlphabet{\mathbb}{AMSb}

 \newcommand{\beqn}{\begin{eqnarray}}
 \newcommand{\eeqn}{\end{eqnarray}}
 \newcommand{\be}{\begin{equation}}
 \newcommand{\ee}{\end{equation}}
  
  \newcommand{\bcor}{\begin{corollary}}
 \newcommand{\ecor}{\end{corollary}}
 
 \newcommand{\bpr}{\begin{proof}}
 \newcommand{\epr}{\end{proof}}
 
 \newcommand{\ba}{\begin{array}}
 \newcommand{\ea}{\end{array}}

 \newcommand{\pa}{\partial}
 
  \newcommand{\ci}{\cite}
 \newcommand{\ds}{\displaystyle}
 \newcommand{\la}{\label}
 \newcommand{\rIm}{{\rm Im\5}}
 \newcommand{\rRe}{{\rm Re\5}}
 \newcommand{\fr}{\frac}

\newcommand{\ov}{\overline}
\newcommand{\rot}{{\rm rot\5}}
\newcommand{\dv}{{\rm div\5}}

\newcommand{\ti}{\tilde}

\newcommand{\cF}{{\cal F}}

\newcommand{\bA}{{\bf A}}

\newcommand{\bba}{{\bf a}}
\newcommand{\bb}{{\bf b}}
\newcommand{\bbc}{{\bf c}}

\newcommand{\bz}{{\bf z}}

\newcommand{\bn}{{\bf n}}
\newcommand{\bv}{{\bf v}}
\newcommand{\bV}{{\bf V}}
\newcommand{\cE}{{\mathbb E}}

\newcommand{\bj}{{\bf j}}

\newcommand{\cm}{{\rm m}}

\newcommand{\cH}{{\cal H}}

\newcommand{\cM}{{\cal M}}
\newcommand{\bS}{{\bf S}}

\newcommand{\cX}{{\mathbb X}}
\newcommand{\cY}{{\mathbb Y}}

\newcommand{\bP}{{\bf P}}
\newcommand{\cP}{{\cal P}}

\newcommand{\bX}{{\bf X}}

\newcommand{\bY}{{\bf Y}}
\newcommand{\bZ}{{\bf Z}}

\newcommand{\ve}{\varepsilon}
\newcommand{\vp}{\varphi}
\newcommand{\De}{\Delta}
\newcommand{\de}{\delta}

\newcommand{\al}{\alpha}
\newcommand{\ga}{\gamma}
\newcommand{\Ga}{\Gamma}
\newcommand{\vka}{\varkappa}

\newcommand{\si}{\sigma}
\newcommand{\om}{\omega}
\newcommand{\Om}{\Omega}
\newcommand{\na}{\nabla}
\newcommand{\Si}{\Sigma}
\newcommand{\lam}{\lambda}

\newcommand{\bPi}{{\bm{\Pi}}}

\newcommand{\Spec}{{\rm Spec\5}}
\newcommand{\5}{{\hspace{0.5mm}}}

\newcommand{\R}{\mathbb{R}}

\newcommand{\C}{\mathbb{C}}

\newtheorem{theorem}{Theorem}[section]

\renewcommand{\thetheorem}{\arabic{section}.\arabic{theorem}}
\newtheorem{defin}[theorem]{Definition}

\newtheorem{lemma}[theorem]{Lemma}
\newtheorem{remark}[theorem]{Remark}
\newtheorem{remarks}[theorem]{Remarks}
\newtheorem{corollary}[theorem]{Corollary}
\newtheorem{pro}[theorem]{Proposition}
\newtheorem{example}[theorem]{Example}

\newcommand{\bp}{\begin{pro}}
\newcommand{\ep}{\end{pro}}
\newcommand{\bl}{\begin{lemma}}
\newcommand{\el}{\end{lemma}}
\newcommand{\bc}{\begin{corollary}}
\newcommand{\ec}{\end{corollary}}

\newcommand{\bd}{\begin{defin}}
\newcommand{\ed}{\end{defin}}
\newcommand{\br}{\begin{remark}}
\newcommand{\er}{\end{remark}}
\newcommand{\brs}{\begin{remarks}}
\newcommand{\ers}{\end{remarks}}

\newcommand{\bex}{\begin{example}}
\newcommand{\eex}{\end{example}}

\newcommand{\bce}{\begin{center}}
\newcommand{\ece}{\end{center}}

\begin{document}
\begin{center}

{\huge On parametric   resonance in the laser action}
\footnote{Preprint arXiv:2208.10179 [quant-ph]}
\bigskip\smallskip

 {\Large A. I. Komech
 }
  \smallskip\\ 
{\it
\centerline{Faculty of Mathematics, University of Vienna}
 }

 \centerline{alexander.komech@univie.ac.at}

\end{center}

\setcounter{page}{1}
\thispagestyle{empty}

\begin{abstract}

We consider the selfconsistent semiclassical Maxwell--Schr\"odinger system
for the solid state laser. The system consists of the Maxwell equations
coupled to $N\sim 10^{20}$ Schr\"odinger equations for active molecules. 
The system includes time-periodic pumping and a weak dissipation.
We introduce the corresponding Poincar\'e map $\cP$  and consider  the differential $D\cP(Y^0)$ at suitable ground state $Y^0$.

We conjecture that the {\it stable laser action} is due to the {\it parametric resonance} (PR) which means that the maximal absolute value of the  corresponding multipliers is sufficiently large. The multipliers are defined as  eigenvalues of $D\cP(Y^0)$. The PR makes the stationary state $Y^0$ highly unstable, and we suppose that this instability maintains the {\it coherent laser radiation}. 
We prove that the spectrum $\Spec D\cP(Y^0)$ is approximately symmetric with respect to the unit circle $|\mu|=1$ if the dissipation is sufficiently small.

More detailed results are obtained for 
the   Maxwell--Bloch system which is a finite-dimensional
approximation of the Maxwell--Schr\"odinger system.
The approximation consists of one-mode Maxwell field
coupled to $N$ two-level molecules. 
We calculate the corresponding Poincar\'e map $P$ by successive approximations. 
The key role in 
calculation of the multipliers
is played by the sum of $N$ positive terms arising in the second-order approximation for the total current. This fact can be interpreted as the {\it synchronization} of the main parts of molecular currents in all active molecules, which is provisionally in line with 
the {\it superradiance} and with
the role of {\it stimulated emission} in the laser action. The calculation of the sum relies on probabilistic arguments which is one of main novelties of our approach. Other main novelties are i) the calculation of the differential $DP$  in the ``Hopf representation'' which corresponds to the factorization of the dynamics by the action of the symmetry gauge group, 
ii)  the justification of this representation,
iii) the block structure of the differential,
and iv) the justification of the ``rotating wave approximation'' by a new estimate for the averaging of slow rotations.

The main peculiarity of the eigenvalue problem  is that  $DP$ is a
 matrix of size   $\sim N\times N$ which
 depends on random distribution of active molecules in the 
 resonance cavity. 
The block structure of  $DP$ allows us to  reduce the eigenvalue problem to  polynomial  equation of  degree four. This reduction relies on a novel 
probabilistic arguments.

  \end{abstract}
  
  \smallskip
  

\tableofcontents

\setcounter{equation}{0}
  \section{Introduction}
  
  The article addresses the problems of laser and maser radiation.   
  In the existing approaches, time-periodic pumping is not included in the dynamical equations, so the corresponding resonance features are absent. This fact was one of our main motivations.  
  
  We consider the 
 selfconsistent
 semiclassical Maxwell--Schr\"odinger system
for the solid state laser. The system consists of the Maxwell equations
coupled to $N\sim 10^{20}$ Schr\"odinger equations for active molecules
with one active electron. 
The system includes time-periodic pumping and a weak dissipation.
We introduce the corresponding Poincar\'e map $\cP$  
and consider  its differential $D\cP(Y^0)$ at suitable ground state $Y^0$.
Let 
$\mu$ denote
 the multipliers which are eigenvalues of the 
differential $D\cP(Y^0)$.
We conjecture that the laser action is due to the parametric resonance, when
\be\la{ph}
 \max_{\mu \in \Spec D\cP(Y^0)} |\mu|> 1.
 \ee

For solutions to the linearized dynamical equation at $Y^0$ with the spectrum satisfying  (\ref{ph}),
the components with $|\mu|\le1$ are bounded, while 
the ones with $|\mu|>1$  grow  exponentially  for large times. 
As a result, {\it  almost all} solutions grow  exponentially
 which is in line with the conventional interpretation of the laser action as the appearance of exponentially growing solutions of the corresponding linearized dynamics, see    (\ci[(6.5.12) and (6.6.23)]{N1973}.
This is why we expect that under the condition (\ref{ph}),
 similar growth takes place  in the case of  sufficiently  small  nonlinear coupling (the growth only occurs
   for limited time because of  nonlinear effects and  dissipation).
  
        The conjecture is confirmed by the fact that laser action is possible only 
 for the pumping above a threshold that is quite similar to
 the parametric resonance \ci{A1989}.
       
      \br{\rm We suppose that the overwhelming predominance of a single mode, corresponding to the maximal value of $|\mu|$,  can be  responsible for the  {\it coherent monochromatic laser radiation}. 
      }
      \er

The main goal of the present article is an analysis of properties
of the multipliers and
 development 
of methods for their calculation.
Our main results are as follows.
\smallskip

I. We develop the Hamiltonian formalism for the ideal case
of the Maxwell--Schr\"odinger system without dissipation. Using this formalism and
 the Lyapunov--Poincar\'e theorem \ci[(3.28)]{YS1975},
 we establish the approximate
 symmetry of the spectrum $\Spec D\cP(Y^0)$ with respect to the unit circle $|\mu|=1$ when the dissipation is sufficiently small (Lemma \ref{lsym} and (\ref{LP222})).
 \smallskip

More detailed results (II and III below) are obtained for 
the   Maxwell--Bloch system which is a finite-dimensional
approximation of the Maxwell--Schr\"odinger system.
The approximation consists of one-mode Maxwell field
coupled to $N$ two-level molecules. 
The  system is time-periodic in the case of  periodic pumping.

 II. We
calculate the corresponding Poincar\'e map $P$ by successive approximations.
The key role in 
calculation of the multipliers
 is played by  
 the sum of $N$ positive terms (\ref{abnn})
 arising in the second-order approximation
  for the total current. This fact
can be interpreted as the {\it synchronization} of the main parts of molecular currents
in all active molecules,
which is provisionally 
in line with the role of {\it stimulated emission} in the laser action.

 III. We calculate suitable approximation for the matrix 
 $DP(Y^0)$
 corresponding to material parameters of the ruby laser.
 The main peculiarity of the corresponding eigenvalue problem  is that  $DP(Y^0)$ is a
 matrix of  size 
 $\sim N\times N$ which
 depends on a random distribution of active molecules in the 
 resonance cavity. Using the
 block structure of this matrix, we  reduce the eigenvalue problem to 
  polynomial equation  of degree four.
  \smallskip
  
 Our main novelties are as follows:
 i) calculation of the sum (\ref{probs2}) and
 the reduction
 to  polynomial equation  of degree six 
 relying on Shnirelman's Ergodic Theorem \ci{S1974,S1993}
 and
 probabilistic arguments (\ref{Sig})--(\ref{Sig3}),
  ii) the block structure of the differential (\ref{Mder3}), 
 and iii)
the justification of the  {\it rotating wave approximation}  by
  a new estimate for averaging of slow rotations (Lemma \ref{lA}).

\br\la{ref}
 {\rm 
 The condition (\ref{ph}) alone is not sufficient for the stable laser/maser action:
 to be sure that the energy of the Maxwell field increases indeed, we check that the projections of the corresponding 
 eigenvectors onto the  Maxwell field do not vanish, see (\ref{cph}) and Remark \ref{rM}.
 
 }
 \er

 Let us comment on previous theories of laser radiation.
The today laser/maser theory   exists on the following three levels.
\smallskip\\
{\it Rate equations.} This approach is based on balance equations for the numbers of emitted and absorbed 
photons \ci{AE1975,H1984, N1973,SSL1978,S2010}. 
\smallskip\\
{\it Semiclassical Maxwell--Schr\"odinger theory.}
This theory is based on
 finite-dimensional approximations of
 the Schr\"odinger equations for each active molecule coupled to
  the classical Maxwell equations
\ci{AE1975, AT1955,H1984,N1973,SSL1978,SZ1997,S2010}.
\smallskip\\
{\it Quantum Electrodynamics.} This approach is based on quantized Maxwell
equations in the  Dicke model 
\ci{D1954,H1984, N1973,SSL1978,S2010} and \ci{HL1973a}--\ci{HL2004}.
\smallskip

The theory of the rate equation comes back to famous Einstein's article \ci{Ei1917} developing the theory 
of  {\it stimulated} and {\it spontaneous emission and absorption}.
\smallskip

  The semiclassical approach uses method of slowly varying
  amplitudes \ci[Sections 6.2 and 6.3]{N1973}. 
  The formation of the coherent  monochromatic  
 radiation is explained by an instability in the equation for amplitude of the  Maxwell mode under various hypothesis on the density of population  of active molecules \ci[Sections 6.5 and 6.6]{N1973}.   
  The criterion of the instability is formulated in terms of presence of
  an unstable (nonnegative) root of the corresponding characteristic equation (\ci[(6.5.12) and (6.6.23)]{N1973}). 
     \smallskip

 The quantum theory of laser action 
 relies on quantized Maxwell
equations in the  Dicke--Haken--Lax 
model 
\ci{HL2004}.
This theory resulted in successful explanation of many laser phenomena: 
superradiance, photon echoes, self-induced transparency and others
\ci{G2011,H1984, N1973,SSL1978,SZ1997,S2010}.
 The results by K. Hepp and E. Lieb \ci{HL2004} establish the phase transitions which occur 
 in the thermodynamical limit as $N\to\infty$
 if the coupling constant is sufficiently large. 
 \smallskip

Note that the time-dependent pumping with a resonant frequency plays the crucial role 
sustaining the  population inversion.
At the same time,
 the existing theories 
 \ci{AE1975, AT1955,H1984,M1997,N1973,SSL1978,SZ1997,S2010}
 ignore the time-dependence of the
pumping, modelling the maintenance
 of the population inversion
 by {\it phenomenological constant terms}
 in the corresponding  dynamical  equations;
 see
  \ci[(6.4.11), (6.4.12)]{N1973} for the semiclassical model and
  the beginning of Section 7.3 in \ci{N1973}
   for the quantum model. So, these models do not include an external  periodic pumping, and the resonance features  are not considered.

\smallskip

In concluzion,  the case of quantum Maxwell field will be considered elsewhere.


\setcounter{equation}{0}
 \section{Laser equations} 
 We use the Heaviside--Lorentz units ({\it  unrationalized Gaussian units}), in which  the main physical constants
(electron charge and mass, Planck's constant, and the speed of light in vacuum)
read \ci[p. 221]{W2002}
\be\la{HL}
  e = -4.8  \times 10^{-10} {\rm esu}, \quad \cm = 9.1 \times 10^{-28} {\rm g},\quad  
 \hbar= 1.055 \times 10^{-27} \mbox{\rm erg$\cdot$ s},\qquad c = 3.0\times 10^{10} {\rm cm/s}.
 \ee
{\bf Laser cavity and active molecules.} The laser resonator  is a (usually cylindrical)
 cavity $ V \subset \R ^ 3 $ with metallic walls (ideal conductor) connected with  the output waveguide.
 Active molecules occupy a subregion $V_a\subset V$.
 Typical industrial laser consists  of two cylindrical lamps 
 of identical size with glass walls
 placed at the foci of a metallic cavity $V$ with an elliptic cross-section.
 The {\it active lamp} $V_a$ is filled with a dielectric medium doped with identical
active molecules. The other {\it pumping lamp} is the
discharge light source which produces the pumping field.
 In particular, for  the typical
 cylindrical lamp of length $12\, cm$ and diameter $0.6\, cm$ we have
 \be\la{VV}
 |V_a|= \fr{12\pi0.36}4\approx 3.4\, cm^3.
 \ee
 The active region $V_a$ is filled with a dielectric medium of electrical conductivity $\si$.
For example, the ruby laser is filled with corundum of  electrical conductivity
\be\la{corsi}
\si=10^{-14}Om^{-1}\cdot cm^{-1}=10^{-2}s^{-1}.
\ee
 The active region $V_a$  is filled with active molecules located at the points 
$x_n$ with numbers
$ n = 1, \dots, N \sim 10^{20}$.
 Suppose, for simplicity of notation, that in each active molecule only
one electron is involved in this interaction, and it is subject to an effective static molecular
potential 
\be\la{Phik}
\Phi_n(x)=\Phi(R_n (x-x_n)),\qquad x\in\R^3,\qquad R_n\in SO(3).
\ee
Here $\Phi$ is the potential of the ion (or nucleus) with the total charge $|e|>0$.
Further we consider two different hypothesis on the distribution of 
 the {\it random values} $x_n$ and  $R_n$ with  $n\in\ov N:=(1, \dots, N)$. 
 \smallskip\\
 {\bf H1. Polycrystalline  medium (or "glass medium" \ci{S2012}).} 
  The   random values $x_n\in V$ and  $R_n\in SO(3)$
 are  uniformly distributed and independent, and 
 almost independent from $x_{n'}, R_{n'}$
 with large $|x_n-x_{n'}|$.
   \smallskip\\
 {\bf H2. Crystalline medium.} 
 The  values $x_n\in V$ 
 are  uniformly distributed, while
 $R_n$ do not depend on 
 $n\in\ov N$.
    \smallskip\\
{\bf Maxwell--Schr\"odinger equations.}
Each active molecule is described by the corresponding wave function
$\psi_n(x,t)$,  and in Heaviside--Lorentz units.
Neglecting spin and scalar potential (both can be easily added), 
 the corresponding Maxwell--Schr\"odinger system 
 reads (see
\ci{BT2009,BF1998,CG2004,GNS1995,K2013,K2022,K2019phys,KKumn2020,PS2014,S2006,S2003}
for the equations without dissipation)  
  \be \la {MSm}
\left\{
\ba{rcl}
\fr1{c^2}\ddot \bA (x, t)&=& \De \bA (x, t) - \fr\si{c^2}\dot \bA (x, t) 
+\fr 1c P_s\bj(\cdot,t),\,\,x\in V

\\
\\
i \hbar\dot \psi_n (t)&=&
H_n(t)\psi_n(t),\,\,\,n=1,\dots, N
\ea\right|,
\ee
where $\si>0$ is electrical conductivity of the cavity medium.
We assume the Coulomb gauge \ci{Jackson}
\be\la{CgA}
\dv \bA (x, t) \equiv 0,
\ee
and we denote by
 $P_s$  the orthogonal projection onto divergent-free vector fields  
in the Hilbert space $L^2(V)\otimes\R^3$. Further,
\be\la{EHm}
H_n(t):=
 \fr1{2\cm}D^2(t)  + e\Phi_n (x).
\ee
 Here
 \be\la{DA}
 D(t)=-i\hbar\na-\fr ec [\bA(x,t)+\bA_p(x,t)], 
\ee 
 where $\bA_p(x,t)$ is an external pumping potential, also
 in the Coulomb gauge and time-periodic:
 \be\la{Apdiv}
  \dv\bA_p(x,t)\equiv 0;\qquad \bA_p(x,t+T)=\bA_p(x,t),\,\,\,t>0,
  \ee 
  where $T>0$.
 The 
 current density is defined by 
\be \la {rjim}
 \bj (x, t) =
\fr e\cm
\sum_n
 \rRe [\ov\psi_n(x,t)D(t)\psi_n(x,t)].
\ee
This formula
neglects overlapping of the supports of the wave functions since the distance between 
active molecules is sufficiently large: it is of order $10$ molecular diameters
when the density of active molecules is of order $10^{20}$.

Note that  the pumping 
  field is a solution to the Maxwell equations in the cavity $V$
  excited by the discharge  currents in the pumping lamp.
  
\noindent{\bf Boundary conditions.}
We choose the boundary conditions 
modelling  ideally conducting 
diamagnetic materials (like cooper, silver, gold, etc).
More precisely, we assume that  
\be\la{BCAi}
\bn (x) \times  \bA (x, t) = 0,\quad \bn(x)\cdot\rot\bA(x,t)=0;\quad \psi_n(x,t)= 0,\,\,\,\,
n\in\ov N; \qquad x \in \Ga,
\ee
where 
$ \bn (x) $ is the outward normal to $\Ga$ at a point $ x \in \Ga $.

\br
{\rm
The boundary conditions (\ref{BCAi}) must to be valid for the total Maxwell
field $\bA (x, t) +\bA_p (x, t) $. We require the same boundary conditions for 
$\bA (x, t)$ to separate the {\it own field} $\bA (x, t)$ from the 
pumping field  $\bA_p (x, t)$. In this case
the pumping field  $\bA_p (x, t)$  satisfies the same boundary conditions
as $\bA (x, t)$. 
}
\er

Under the Dirichlet boundary conditions for the wave functions,
the total charge of each active molecule is constant, i.e.,
\be\la{qcons}
\int_V|\psi_n(x,t)|^2dx=1,\qquad t>0;\qquad n\in\ov N.
\ee
\noindent{\bf Hamiltonian structure.}
In the {\it ideal case}, when $\si=0$,
the system (\ref {MSm}), under the boundary conditions (\ref{BCAi}),
is formally Hamiltonian, with the
Hamiltonian functional (which is the energy up to a factor)
\be\label{enc}
\cH(\bA,\bPi,\psi,t)=
\fr12[\Vert \fr1c \bm{\Pi}\Vert^2
+\Vert \rot\bA\Vert^2]+\sum_n\langle\psi_n, H_n(\bA,t)\psi_n\rangle,
\ee
where $\psi:=(\psi_1,\dots,\psi_{N})$,
$\Vert\cdot\Vert$ stands for the norm in the real Hilbert space\index{Hilbert space}
$L^2(\R^3)\otimes\R^3$, and
the brackets
$\langle\cdot,\cdot\rangle$ stand for the inner product in
$L^2(\R^3)\otimes\C$.
The  Schr\"odinger operators are defined by (\ref{EHm}), (\ref{DA}):
\be\la{Sope}
H_n(\bA,t):=
\fr 1 {2m} [- i \hbar \na- \ds \frac ec (\bA ( x)+\bA_p ( x,t))]^2
+e\Phi_n(x).
\ee
The system \eqref{MSm} with $\si\ge 0$ can be written in the Hamiltonian form
with variational derivatives,
\be\label{MSH}
\left\{\ba{l}
\fr 1{c^2} \dot\bA(t)=D_\bPi\cH(\bA(t),\bPi(t),\psi(t),t),
\,\,\,
\fr 1{c^2} \dot\bPi(t)=-D_\bA\cH(\bA(t),\bPi(t),\psi(t),t)
-\fr\si{c^2}\dot\bA(x,t)
\\\\
 i\hbar\dot\psi_n(t)=\fr12D_{\psi_n}\cH(\bA(t),\bPi(t),\psi(t),t)=H_n(\bA(t),\psi(t),t)\psi_n(t),
\ea\right|.
\ee
In the last equation, $D_{\psi_n}$ denotes
the variational derivative with respect to 
real and imaginary parts of $\psi_n$, and
the 
factor $\fr12$ is due to this identification.
 The factor $i$ in this case is identified 
with the corresponding skewsymmetric $2\times 2$
matrix. 
\smallskip\\
{\bf Symmetry group.}
 For any $\Theta_n\in\R$, the functions $\bA(x,t)$, $e^{i \Theta_n}\psi_n(x,t)$ are solutions 
of  the system (\ref{MSm}) if $\bA(x,t)$, $\psi_n(x,t)$ is a solution.
In other words, the dynamics (\ref{MSm}) under the boundary conditions (\ref{BCAi}) commutes with the action of the
 {\it symmetry  group} 
$
G:=[U(1)]^{N}
$
with the action
\be\la{actG}
(e^{i\Theta_1},\dots, e^{i\Theta_{N}})(\bA,\psi_1,\dots, \psi_{N})=
(\bA,e^{i\Theta_1}\psi_1,\dots, e^{i\Theta_{N}}\psi_{N}).
\ee

 \setcounter{equation}{0}
\section{Parametric resonance}
Let us denote $L^2=L^2(\R^3)$.
  The system (\ref{MSH}) can be written as the dynamical 
 equation on the phaase space 
  $\cY=[L^2(\R)\otimes\R^3]^2  \oplus  [L^2]^{2N}$:
\be\la{XF2}
\dot Y(t)=\cF(Y(t),t),
\qquad
Y(t):=(\bA(t),\bPi(t),\psi(t)).
\ee
This system  is   $T$-periodic by (\ref{Apdiv}), i.e.,
\be\la{XFp23}
\cF(Y,t+T)\equiv \cF(Y,t).
\ee
The corresponding 
   Poincar\'e map  is defined by
  \be\la{Po22}
  \cP:Y(0)\mapsto Y(T).
  \ee
  Let $Y^0(t)$ be a ``stationary state"
of the system (\ref{XF2}) with the zero pumping $\bA_p(x,t)=0$, i.e., a solution of type 
\be\la{station}
Y^0(t)=(\bA,\bPi, \psi(0)e^{-i\om t}),
\ee
where $\om\in\R$.
The following definition is relevant for the system  (\ref{XF2}) with
sufficiently small pumping. We will denote by $Y^0$ every value
of the function (\ref{station}).

\bd\la{dph2}
{\rm
The parametric resonance  takes place for $Y^0$ when the condition
 (\ref{ph}) holds.
 }
 \ed

 \setcounter{equation}{0}
\section{Symmetry of spectrum}
 
In the 
{\it  ideal case}, when  $\si=0$, the dissipation is absent, and
 the  system (\ref{XF2}) is Hamiltonian by (\ref{MSH}), 
 so it   can be written as
  \be\la{MB31cH}
 \dot Y(t)=JD_Y \cH(Y(t)),t),
\ee
where
   $J$ is a skewadjoint operator.
      We can make
      $J^2=-1$
   choosing  suitable units wiht $c=2\hbar=1$
   (note that the multipliers do not depend on the choice of units). 
  Denote by $\cP:Y(0)\mapsto Y(T)$ the corresponding Poincar\'e map.

 \bl\la{lsym}{\rm (Lyapunov--Poincar\'e theorem)}
 In the ideal case,
 the symmetry holds
 \be\la{LP22}
  \Spec D\cP(Y^0)= R (\Spec D\cP(Y^0)),
 \ee
 where  $R\mu={\ov\mu}^{-1}$ is the inversion, which is the reflection in the unit circle $|\mu|=1$. 
 \el
 \bpr
  The  differential $D\cP(Y^0)$
 admits the representation 
 \be\la{DPr}
 D\cP(Y^0)\bY (0)=\bY(T),
 \ee
  where
 $\bY(t)$ is the solution
 to the linearized equation  
 \be\la{Ut}
 \dot \bY(t)= K(t) \bY(t), \qquad K(t)=D_Y \cF(Y^0_*(t),t),
 \ee
  and $Y^0_*(t)$ is the solution to (\ref{MB31cH}) with 
  initial condition $Y^0_*(0)=Y^0$. 
  However, (\ref{MB31cH}) implies that
 $$
 \cF(Y,t)=JD_Y \cH(Y,t),
 $$
  and therefore,
 \be\la{AH}
 K(t)=JD^2_Y \cH(Y^0_*(t),t)=JA(t),
 \ee
 where $A(t)$ is the selfadjoint operator.
 Denote the map $U(t):\bY(0)\to \bY(t)$, so 
 $D\cP(Y^0)=U(T)$ by (\ref{DPr}).
 Finally,
 the Lyapunov--Poincar\'e arguments \ci[(3.28)]{YS1975}
 imply that $\Spec U^*=\Spec U^{-1}$, where $U=U(T)$.
 Hence,
the symmetry
(\ref{LP22}) is proved. 
\epr
 
 Recall the arguments \ci[(3.28)]{YS1975}.
  We have
  $\dot U(t)=JA(t)U(t)$, so
 differentiating  $W(t):=U^*(t)JU(t)$, we obtain
 \be\la{UY}
 \dot W=\dot U^*JU+U^*J\dot U= U^*A(-J^2)U+U^*J^2 A U=0
 \ee
 since $J^2=-1$ commutes with $A$.
Therefore, $U^*(t)JU(t)= J$, and hence $U^*(t)=JU^{-1}(t)J^{-1}$.
 This similarity implies that the spectra  of $U^*(t)$ and $U^{-1}(t)$
 coincide.

 \br
{\rm
The symmetry (\ref{LP22}) can be considered as a nonlinear version of the 
Lyapunov--Poincar\'e theorem, which plays the key role in M.G. Krein's theory of parametric resonance \ci{DK1974,YS1975}.
The ground for this extension is the Hamiltonian structure of the dynamics (\ref{MB31cH}).

}
\er

 Thus, in the ideal case,
 the set of multipliers is symmetric with respect to the unit circle.
 In general case, when the dissipation is included, the symmetry (\ref{LP22}) is broken, though 
 is holds approximately for sufficiently small $\si>0$:
  \be\la{LP222}
 \Spec D\cP(Y^0)\approx R (\Spec D\cP(Y^0)).
 \ee 
 Accordingly, the parametric resonance (\ref{ph}) is 
  very likely
  in this case.
The required smallness defines
 the 
 pumping/damping {\it threshold}.
  \br\la{rred}
  \rm
 Similar symmetry also holds for the case of the Maxwell--Bloch equations (\ref{MB31}) 
  which makes the parametric resonance
  (\ref{ph}) {\it very plausible}.

\er

\setcounter{equation}{0}
\section{Dipole approximation} 
The theory
admits a  significant  simplification in the case when 
the wavelength of the pumping  
is large with respect to the size of active molecules $D$.
\\
{\bf H3.
The dipole radiation condition.} We will assume that
 \be\la{dipap}
 \lam=cT=2\pi c/\Om_p\gg D,
\ee
where $\Om_p=2\pi/T$ is the frequency of the pumping,
\bex
{\rm
This condition holds for the ruby laser since the wavelength $\lam\approx 7\times 10^{-5}\,cm$ while the size of the chromium molecule is
 \be\la{DD}
 D\approx2.5 pm=2.5\times 10^{-10}cm.  
\ee
}
\eex
Taking into account the condition (\ref{dipap}), we approximate the system 
(\ref{MSm}) by
\be \la {MSrd}
\!\!\!\!\!\!
\left\{\ba{rcl}
\fr1{c^2}\ddot \bA (x, t) &=& \De \bA (x, t) 
- \fr\si c \dot \bA (x, t)+\fr1{c} P_s\hat\bj (x, t)
 \\
\\
 i \hbar\dot \psi_n (x, t)&=&
 \hat H_n(t)\psi_n(t),\,\,\,n=1,\dots, N
\ea\right|, \quad x \in V.
\ee
Here 
\be\la{EHmd}
\hat H_n(t):=
 \fr 1{2\cm}D_n^2(t)  +e\Phi_n (x),\qquad
 D_n(t)=-i\hbar\na-\fr ec [\bA(x_,t)+\bA_p(x_n,t)],
 \ee 
 The current (\ref{rjim}) in the dipole approximation
becomes 
\be \la {rjr}
\bj (x, t) \approx \fr e\cm \rRe \sum_n \ov \psi_n (x, t) D_n(t) \psi_n (x, t).
\ee
 The system (\ref{MSrd}) with $\si=0$ is also Hamiltonian with the Hamilton functional 
\be\label{enc2}
\hat\cH(\bA,\bPi,\psi,t)=
\fr12[\Vert \fr1c \bm{\Pi}\Vert^2
+\Vert \rot\bA\Vert^2]+\fr12\sum_n\langle\psi_n, \hat H_n(\bA,t)\psi_n\rangle,
\ee
where
\be\la{EHmd2}
\hat H_n(\bA,t):=
 \fr 1{2\cm}D_n^2(\bA,t)  +e\Phi_n (x),\qquad
 D_n(\bA,t)=-i\hbar\na-\fr ec [\bA(x)+\bA_p(x_n,t)],
 \ee

\setcounter{equation}{0}
\section{Gauge transform}

The dipole approximation (\ref{MSrd}) allows us to 
apply the gauge transform \ci[(5.1.15)]{SZ1997}:
\be\la{gatr} 
\psi_n(x,t)=e^{i\chi_n(x,t)}\ti\psi_n(x,t),\qquad \chi_n(x,t):=
\fr e{\hbar c} (x-x_n)[\bA(x_n,t)+\bA_p(x_n,t)].
\ee
Now the
dipole approximation (\ref{MSrd}) becomes  
\be \la {MSrdg}
\!\!\!\!\!\!
\left\{\ba{rcl}
\fr1{c^2}\ddot \bA (x, t) &=& \De \bA (x, t) 
- \fr\si c \dot \bA (x, t)+\fr1{c} P\bj (x, t)
 \\
\\
 i \hbar\dot {\ti\psi}_n (x, t)&=&
 \ti H_n(t)\ti \psi_n(t),\,\,\,n=1,\dots, N
\ea\right|, \quad x \in V.
\ee
The Schr\"odinger operator now reads
\be\la{rjr2} 
 \ti H_n(t)= -\fr 1{2\cm}(-i\hbar\na -\fr ec[\bA(x,t)-\bA(x_n,t)])  +e\Phi_n (x) +\hbar\dot\chi_n(x,t)\approx
 -\fr{\hbar^2}{2m}\De+e\Phi_n (x)+\hbar\dot\chi_n(x,t),
 \ee
 where we used
 the condition (\ref{dipap}).
  The current (\ref{rjr}) now reads as
\be \la {rjrb}
 \bj (x, t) \approx \fr e\cm \rRe \sum_n \ov {\ti\psi}_n (x, t) (-i\hbar\na-\fr ec[\bA(x,t)-\bA(x_n,t)]) \ti\psi_n (x, t)\approx
 \fr e\cm \rRe \sum_n \ov {\ti\psi}_n (x, t) (-i\hbar\na) \ti\psi_n (x, t).
\ee

   \setcounter{equation}{0}
\section{Decoupled system and  eigenfunction expansions}
{\bf Decoupled system.}
The system (\ref{XF2}) 
is a
 weak perturbation of the corresponding ``decoupled''  system
 obtained by neglecting the interaction terms with the coupling constant 
 $e/c$:
 \be \la {MSmf}
\left\{
\ba{rcl}
\dot  \bA (x, t)&=&\bPi(x,t)
\\
\\
\fr1{c^2}\dot \bA (x, t)&=& \De \bA (x, t) - \fr\si{c^2}\dot \bA (x, t) 
\\
\\
i \hbar\dot \psi_n (t)&=&
H_n^0\psi_n(t),\,\,\,n\in\ov N
\ea\right|,
\ee
 where
 \be\la{EHmf}
H_n^0:=
 -\fr{\hbar^2}{2\cm}\De  + e\Phi_n (x).
%
\ee
{\bf Schr\"odinger modes.}
The decoupled system (\ref{MSmf}) can be solved explicitly
using the orthonormal eigenfunctions of 
the Schr\"odinger operators $H_n^0$
under the Dirichlet boundary conditions (\ref{BCAi}):
\be\la{eif}
 H_n^0\vp_{n,l}(x)=\om_{n,l}\vp_{n,l}(x), \qquad l=1,2,\dots
\ee
Hence, the system (\ref{XF2}) can be analyzed by perturbation methods.
The eigenfunctions $\vp_{nl}$  differ by rotations (\ref{Phik}), while
 the eigenvalues $\om_{n,l}=\om_l$  do not depend on $n$ with a great accuracy.
The decoupled system (\ref{XF2}) admits the
``stationary" solutions 
\be\la{stsol2}
Z^0(t)\equiv (0, 0,(e^{-i\om_{l(n)} t }\vp_{n,l(n)}:n\in\ov N)),
\ee
where $l(n)=1,2,\dots$.
\smallskip\\
{\bf Maxwell modes.} 
The first equation of the decoupled system (\ref{MSmf}) can be solved using
 the orthonormal 
 eigenmodes 
\be \la {eigA}
\left\{
\ba{rcl}
 \De \, \bX_\nu (x)& =& - \fr{\Om_\nu ^ 2}{c^2} \bX_
 \nu (x), \quad \dv \bX_
 \nu (x) = 0, \quad x \in V
 \\
 \\
 \bn (x) \times \bX_
\nu(x) &=& 0, \qquad\,\,\,
 \bn (x) \cdot \rot  \bX_
\nu (x) = 0,  \quad x \in \pa V
\ea\right|,
\ee
and  $\Om_\nu>0$.

\bex\la{ex1}
{\rm
The eigenmodes  can be calculated explicitly for
the rectangular cuboid 
$V=[0,l_1]\times[0,l_2]\times[0,l_3]$: 
\be\la{eigcubo}
\bX_k=C
\left(
\ba{c}
a_k^1\cos \fr{k_1\pi x_1}{l_1}\sin \fr{k_2\pi x_2}{l_2}
\sin \fr{k_3\pi x_3}{l_3}\\
a_k^2\sin \fr{k_1\pi x_1}{l_1}\cos \fr{k_2\pi x_2}{l_2}
\sin \fr{k_3\pi x_3}{l_3}\\
a_k^3\sin \fr{k_1\pi x_1}{l_1}\sin \fr{k_2\pi x_2}{l_2}
\cos \fr{k_3\pi x_3}{l_3}
\ea
\right),\qquad C=\sqrt{\fr{8}{l_1l_2l_3}},\qquad k=(k_1,k_2,k_3),\quad
k_j=1,2,\dots,
\ee
where the amplitudes $a_k=(a_k^1,a_k^2,a_k^3)$ are unit vectors,
orthogonal to the wave vectors $k$.
In particular, consider the cuboid $V$ 
with 
 dimensions 
 $l_1=12cm$, $l_2=l_3=2cm$.
  Note that   the random points $x_n\in V_a$ are distributed uniformly, so 
  ``the dispersion''
 of the ``random value" $\bX_k(x_n)$ is
 \be\la{modX0}
D(\bX(x_n))=
\fr1{|V_a|}\int_{V_a} \bX_k^2(x)dx
\approx
\fr1{|V_a|}\fr{|V_a|}{|V|}
=\fr{1}{|V|}\approx 0.02.
 \ee
Accodingly, the mean value is 
  \be\la{modX}
  |\bX(x_n)|\approx0.14.
\ee
  
  }
  \eex
  
\noindent{\bf Eigenfunction expansions.}
 We will expand $\ti\psi_n(\cdot,t)$ in the orthonormal eigenfunctions 
 (\ref{eif}):
\be\la{exp0inf}
\ti\psi_n(x,t)= \sum_{l=1}^\infty c_{n,l}(t)e^{-i\om_l t}\vp_{n,l}(x)
=e^{-\fr i\hbar H_n^0 t} \phi_n(t),\qquad \phi_n(t):=\sum_{l=1}^\infty c_{n,l}(t)\vp_{n,l}(x).
\ee
Now each Schr\"odinger equation in (\ref{MSrdg}) 
with  $n\in\ov N$
is equivalent to the system
\be\la{psik5}
 \dot c_{n,l}(t)=-i\sum_{l'} e^{-i\om_{l,l'} t}
 \langle \dot\chi_n(x,t) \vp_{n,l'}(x),\vp_{n,l}(x)\rangle c_{n,l'}(t),\quad l\ge 1;\quad \om_{l,l'}=\om_{l'}-\om_{l}.
 \ee
Similarly,  the Maxwell field can be expanded in the orthonormal 
eigenmodes (\ref{eigA}): 
\be \la {maxA}
  \bA (x, t) = \sum_\nu a_\nu (t) \bX_\nu (x).
\ee
 Now the Maxwell equation in  (\ref{MSrdg}) is equivalent to the system
\be\la{Mexp}
\ddot a_\nu(t)+c\si \dot a_\nu(t)+\Om_\nu^2a_\nu(t)=c\langle\bj(t),\bX_\nu\rangle,
\qquad \nu=1,\dots
\ee

\setcounter{equation}{0} 
\section{Resonance condition} 
The resonance condition means that
 \be\la{res122}
\De=\Om_p=\Om_{\ov\nu},\qquad \De:=\om_a-\om_b,
\ee
where $\om_b<\om_a$ are some fixed eigenvalues (\ref{eif}), $\Om_{\ov\nu}$
is one of the Maxwell eigenfrequencies (\ref{eigA}), and $\Om_p=2\pi/T$
is the frequency of the pumping (\ref{Apdiv}).

We can renumerate the eigenfunctions  $\vp_{n,l}$ in such a way that  $b=1$ and $a=2$. 
Then the   decoupled system (\ref{MSmf}) admits solutions (\ref{stsol2}) with $l(n)\equiv 1$:
\be\la{stsol33}
Y^0(t)\equiv (0, 0,(e^{-i\om_{1} t }\vp_{n,1}:n\in\ov N)),
\ee
Let us call as  ``ground states" all values of these solutions:
in particular,
\be\la{stsol3}
Y^0= (0,0, (\vp_{n,1}: n\in\ov N)).
\ee
The perturbation theory allows us to calculate 
solutions to the coupled system  (\ref{MSm}) 
with initial states
close to (\ref{stsol3}). We expect that these solutions  are 
rapidly growing due to the resonance condition (\ref{res122})
and the parametric resonance (\ref{ph}).

 \setcounter{equation}{0}
\section{The harmonic pumping}
We consider the Maxwell--Schr\"odinger
system (\ref{MSm}) with the   pumping $\bA_p(x,t)$, which is 
switched on at $t=0$ and is 
      monochromatic,
i.e., 
\be \la {Ap}
\bA_p(x,t)=
\bba_p (x)\sin\Om_p t,
\ee
where $\Om_p$ is the pumping frequency.
For example, for the ruby laser the wavelength and the resonant pumping 
frequency are 
\be\la{lamOm}
\lam=694.3 nm=694.3 \times 10^{-9}m\approx 7\times 10^{-5}\,cm, 
\qquad \Om_p=\fr{2\pi c}\lam=\fr{19\times 10^{10}}{7\times 10^{-5}}\approx 3\times 10^{15} s^{-1}.
\ee
{\bf The magnitude of pumping.} 
The pumping field 
in a neighborhood of any active molecule
is approximately a plane wave 
\be\la{pw}
\bA_p(x,t)
\approx \bba_p\sin(\Om_p t-kx),\qquad |k|=|\Om_p|/c,
\ee
or a mixture of such waves.
Let us estimate the pumping field in a typical
 cylindrical cavity of length $12\,cm$  and diameter $0.6\, cm$ 
 illuminated by the pumping lamp of the typical power  $W=1\, Kwt=10^{10}\,erg/s$. The side surface $S$ of the cavity  is about $12\pi0.6\, cm^2\approx23 cm^2 $.
The intensity $I$ of the plane wave (\ref{pw}) is  $\fr{ck^2|\bba_p|^2}{2}$ 
$\fr{erg/s} {cm^2}$, so
\be\la{Kwt}
W=IS=\fr{ck^2|\bba_p|^2}{2} 12\pi0.6=10^{10}.
\ee
Substituting $ |k|=|\Om_p|/c$, we find
\be\la{Kwt2}
\fr{\Om_p^2|\bba_p|^2}{2c} 23= 10^{10}.
\ee
For example, for the ruby laser with frequency (\ref{lamOm}), we have 
\be\la{Kwt30}
|\bba_p|^2\approx\fr{c}{\Om_p^2}\times10^{9}=0.3\times 10^{-11}.
\ee
 Finally, 
 \be\la{Kwt3a}
|\bba_p|\approx  1.7\times10^{-6}\,\,esu/cm.
\ee
\noindent{\bf H4. Distribution of pumping.} 
We assume that  i) $\bba_p(x_n)$ is the random vector
distributed uniformly on the sphere
$
|\bba_p(x)|= a_p
= 1.7\times10^{-6},
$
and ii) $\bba_p(x_n)$ and $\bba_p(x_{n'})$ 
are  almost independent  at large distances $|x_n-x_{n'}|$.

 \setcounter{equation}{0}
\section{The Maxwell--Bloch equations}
Now we consider the finite-dimensional 
Faedo--Galerkin approximation  of the Maxwell--Schr\"odinger system (\ref{XF2}).
Namely,
we will consider  solutions
\be\la{solMB}
\bA(x,t)=a(t)\bX_{\ov\nu}(x),\qquad
\ti\psi_n(x,t)=c_{n,1}(t)\vp_{n,1}(x)+c_{n,2}(t)\vp_{n,2}(x)
\ee
which satisfy the system
consisting of   two equations (\ref{psik5}) with $n=1$ and $n=2$
for each $n\in\ov N$,
 and one equation of (\ref{Mexp}) with $\nu=\ov\nu$.
Below, we 
will write $a(t)$, $\Om(\tau)$ and $\bX(x)$ instead of $a_{\ov\nu}(t)$, $\Om_{\ov\nu}(\tau)$ and $\bX_{\ov\nu}(x)$ respectively.
Thus, the approximation of the  Maxwell--Schr\"odinger equations reads as
the Maxwell--Bloch equations
\beqn\la{MB1}
\ddot a(t)+c\si \dot a(t)+\Om^2a(t)=c\langle\bj(t),\bX\rangle,
\qquad\qquad\qquad\qquad\qquad\qquad\qquad\qquad
\\
\nonumber\\
\left\{\ba{rcl}
 \dot c_{n,1}(t)&=&-i \sum_{l'=1}^2e^{-i\om_{1,l'} t}
 \langle \dot\chi_n(x,t) \vp_{n,l'}(x),\vp_{n,1}(x)\rangle c_{n,l'}(t)
\\
\\
\dot c_{n,2}(t)&=&-i \sum_{l'=1}^2 e^{-i\om_{2,l'} t}
 \langle \dot\chi_n(x,t) \vp_{n,l'}(x),\vp_{n,2}(x)\rangle c_{n,l'}(t)
 \ea\right|,\qquad n\in\ov N\la{MB12}.
 \eeqn

\setcounter{equation}{0}
\section{Molecular currents}

By  (\ref{dipap}),  the matrix entries in (\ref{MB12}) 
read as
\beqn\la{maen}
b_n^{ll'}(t)&:=& \langle \dot\chi_n(x,t) \vp_{n,l'}(x),\vp_{n,l}(x)\rangle\approx\fr e{\hbar c}
[\dot\bA(x_n,t)+\dot\bA_p(x_n,t)]
\langle (x-x_n)\vp_{n,l'}(x),\vp_{n,l}(x)\rangle
\nonumber\\
\nonumber\\
&=&\fr e{\hbar c}
[\dot a_1(t)\bX(x_n)+\bba_p(x_n)\Om_p\cos\Om_pt]
\bP_n^{ll'}
\nonumber\\
\nonumber\\
&=&\fr {\bP_n^{ll'}}{\hbar c}
[\dot a_1(t)\bX(x_n)+\bba_p(x_n)\Om_p\cos\Om_pt],\qquad l,l'=1,2,
\eeqn
where $\bP_n^{ll'}$ are the dipole momenta
 \be\la{Pk}
 \bP_n^{ll'}:=e\langle\vp_{n,l'}(x),(x-x_n)\vp_{n,l}(x)\rangle.
 \ee
 These momenta are real vectors
since we have chosen real  eigenfunctions $\vp_{n,l}(x)$.

 \br\la{rPP}
{\rm
 In the case of the ruby laser, the matrix of type (\ref{Pk}) 
corresponds to the dipole momenta of  chromium ions Cr$^{3+}$.
For definiteness of calculations, we accept 
\be\la{Pval}
|\bP|  \approx 4\,Debye=4\times 10^{-18} esu\cdot cm,
\ee
which is 
the {\it permanent dipole moment} of  chromium oxide molecules
CrO; see \ci[p. 10]{DH1992}.
However,  actual magnitude of the matrix elements (\ref{Pk}) in the case of the ruby laser
must be 
revised.

}
\er

 From now on, we will assume that the potential $\Phi(x)$ is an even function, i.e.,
 \be\la{Peven}
 \Phi(-x)=\Phi(x),\qquad x\in\R^3.
 \ee
 In this case, all eigenfunctions  $\vp_{n,l}$ can be chosen odd or even, and hence,
 \be\la{Pdia} 
 \bP_n^{11}=\bP_n^{22}=0.
 \ee
 Therefore, the 
  system (\ref{MB12}) reads as
 \be\la{Blo}
\left\{\ba{rcl}
 \dot c_{n,1}(t)&=&-i b_n(t) e^{-i\De t}
  c_{n,2}(t)
\\
\\
\dot c_{n,2}(t)&=&-i  b_n(t) e^{i\De t}
  c_{n,1}(t)
 \ea\right|, \qquad n\in\ov N,
 \ee
 where we denote
 \be\la{at}
 b_n(t):=
 \fr1{\hbar c}
[\dot a_1(t)\bP_n \bX_1(x_n)+\bP_n\bba_p(x_n)\Om_p\cos\Om_pt],\qquad
\bP_n:=\bP_n^{12}.
 \ee
The matrix of this system is skew-symmetric, 
which corresponds to the  electron 
charge conservation in each active molecule i.e., 
\be\la{qcons2}
| c_{n,1}(t)|^2+| c_{n,2}(t)|^2=1,\qquad t>0,\quad n\in\ov N.
\ee

 We now 
 calculate the current (\ref{rjrb}) corresponding to the wave functions (\ref{solMB}),
 and the right hand side of (\ref{Mexp})
 with $\bX_\nu=\bX$: taking into account the dipole approximation
 (\ref{dipap}), we obtain that
 \beqn\la{maxrk}
\langle\bj(t),\bX\rangle\!\!\!\!&\!\!\!\!\approx \!\!\!\!&\!\!\!\! \fr {e\hbar}{\cm}
\sum_n
 \bX(x_n)
\rIm\Big\{\ov c_{n,1}(t) c_{n,2}(t)  e^{-i\De t}  \!\!\int\! \ov\vp_{n,1}(x)\na\vp_{n,2}(x)dx  
+
\ov c_{n,2}(t)  c_{n,1}(t)e^{i\De t}\!\!\int\! \ov\vp_{n,2}(x)\na\vp_{n,1}(x)dx \Big\}
\nonumber\\
\nonumber\\
\!\!\!\!&\!\!\!\!=\!\!\!\!&\!\!\!\! 2 \De\sum_n\bP_n  \bX(x_n)
\rIm\Big\{\ov c_{n,1}(t) c_{n,2}(t)  e^{-i\De t} \Big\},
\eeqn
since
\be\la{del}
\int \ov \vp_{n,1} \na \vp_{n,2} dx=\fr{\De\cm}\hbar 
\int \ov \vp_{n,1} x \vp_{n,2} dx=\fr{\De\cm}\hbar 
\fr{\bP_n}e
\ee
by \ci[(44.20)]{Schiff1955}. 

\setcounter{equation}{0}
\section{Time scaling}

Now the Maxwell--Bloch equations (\ref{MB1})--(\ref{MB12}) become
\beqn\la{MB2}
  \ddot a(t)+\Om^2a(t)+c\si \dot a(t)=j(t):=
2 c\De\sum_n\bP_n \bX(x_n)
\rIm\Big\{\ov c_{n,1}(t) c_{n,2}(t)  e^{-i\De t}\Big\},\qquad\qquad
\\
\nonumber\\
\left\{\ba{l}
 \dot c_{n,1}(t)=-i b_n(t) e^{-i\De t}
  c_{n,2}(t)
\\
\\
\dot c_{n,2}(t)=-i  b_n(t) e^{i\De t}
  c_{n,1}(t)
 \ea\right|, \,\,\, n\in\ov N.\qquad\qquad\qquad\qquad\qquad\qquad
 \la{MB22}
 \eeqn
Let us transform the system (\ref{MB2}) to the natural  time scale
\be\la{tau}
\tau:=\Om_pt\approx
3\times10^{15}t, 
\ee
in which the period of pumping
is $2\pi$.
Below, we denote any function $f(t)$ in the units of time (\ref{tau})
also as $f(\tau)$. In particular,
  (\ref{at}) becomes
\be\la{at2}
 b_n(\tau)=\fr{\Om_p}{\hbar c}
[\pa_\tau a(\tau)\bP_n\bX(x_n)+\bP_n\bba_p(x_n)\cos\tau].
 \ee
Now the Maxwell--Bloch equations (\ref{MB2}),  (\ref{MB22}) read as
\be\la{MB2u}
\pa_\tau^2 a(\tau)+\si_1\pa_\tau a(\tau)+\fr{\Om^2}{\Om_p^2}a(\tau)=
j(\tau),\qquad
\left\{\ba{rcl}
 \pa_\tau c_{n,1}(\tau)=-i \fr {b_n(\tau)}{\Om_p}  e^{-i\fr\De{\Om_p} \tau}
  c_{n,2}(\tau)
\\
\\
\pa_\tau c_{n,2}(\tau)=-i  \fr {b_n(\tau)} {\Om_p}   e^{i\fr\De{\Om_p} \tau}
  c_{n,1}(\tau)
 \ea\right|,\,\,\,n\in\ov N
  \ee
 where
 \be\la{epsjj}
  j(\tau):=\fr{2 c\De}{\Om_p^2}\sum_n\bP_n  \bX(x_n)
\rIm\Big\{\ov c_{n,1}(\tau) c_{n,2}(\tau)  e^{-i\fr\De{\Om_p}\tau}\Big\}
\ee
and
\be\la{epsj}
  \si_1:=\fr{c\si}{\Om_p}\approx\si10^{-5}\approx 10^{-7}.
 \ee
Let us denote the real constants
\be\la{albege}
 \al_n=\fr{2 c}{\Om_p}\bP_n  \bX(x_n),\qquad
 \beta_n=\ds\fr{\bP_n\bX(x_n)}{\hbar c},\qquad
  \ga_n=\ds\fr{\bP_n\bba_p(x_n)}{\hbar c}.
\ee
Then the system (\ref{MB2u}) becomes
\be\la{MB31}
\ddot a(\tau)+\si_1 \dot a(\tau)+a(\tau)=
j(\tau),\qquad
\left\{\ba{l}
 \dot c_{n,1}(\tau)=-i\om_n(\tau) 
  c_{n,2}(\tau)
\\
\\
\dot c_{n,2}(\tau)=- i\ov\om_n(\tau)  
  c_{n,1}(\tau)
 \ea\right|, \,\,\, n\in\ov N,
  \ee
  where
  we used (\ref{res122}),
   $\dot a:=\pa_\tau a$, etc, and 
  \be\la{jbk}
 j(\tau)=\sum_n\al_n
\rIm\Big\{\ov c_{n,1}(\tau)c_{n,2}(\tau)  e^{-i\tau}\Big\},
\qquad
  \om_n(\tau)=[\beta_n\dot a(\tau)+\ga_n\cos\tau] e^{-i \tau}.
\ee
The matrix of system (\ref{MB31}) is skew-adjoint, and the charge conservation  holds: similarly to
(\ref{qcons2}), 
\be\la{Blcc}
|c_{n,1}(\tau)|^2+|c_{n,2}(\tau)|^2=1,\qquad \tau\in\R.
\ee

 All coefficients 
 $\al_n$, $\beta_n$ and $\ga_n$ are small. In particular, for the 
 ruby laser with the
 chromium dipole  moment (\ref{Pval}), 
 and the pumping frequency (\ref{lamOm}),
 we have
 \be\la{intes}
\fr{c|\bP_n|}{\Om_p}\sim
\fr{12\times 10^{10}10^{-18}}{3\times10^{15}}\sim 4\times 10^{-23},\qquad 
\fr{|\bP_n|}{\hbar c}\sim\fr{4\times10^{-18}}{10^{-27}3\times 10^{10}}\sim 1.3\times10^{-1}.
\ee
Now 
(\ref{modX}) and (\ref{Kwt3a}) imply that 
\be\la{abg}
|\al_n|\sim 10^{-23},\qquad |\beta_n|\sim 0.2 \times10^{-2},\qquad
|\ga_n|\sim0.7\times10^{-7}.
\ee
In all calculations below we will consider these values. 
\br\la{rH}
\rm 
By our hypotheses {\bf H1}--{\bf H4}, 
the parameters $\al_n,\beta_n,\ga_n$
and $\al_{n'},\beta_{n'},\ga_{n'}$
 are almost independent for sufficiently large distances $|x_{n'}-x_n|$. 

\er

   \setcounter{equation}{0}
\section{Parametric resonance for the Maxwell--Bloch equations}
Let us reformulate the concept of 
parametric resonance (\ref{ph}) for the system (\ref{MB31}).
In the case of zero pumping $\bba_p=0$,
this system  admits ``ground states", which are stationary solutions corresponding to 
 (\ref{stsol33}):
\be\la{stsol}
X^0(t)\equiv X^0:=(a^0,b^0, (c_n^0:n\in\ov N))
\ee
 with
\be\la{zsol}
a^0=b^0=0; \qquad c_n^0=(e^{i\Theta_n},0),\quad n\in\ov N.
\ee
The solutions
describe the laser  dynamics
before the pumping is switched on.
Let us write the system
  (\ref{MB31})
  as
   \be\la{MB31c}
   \dot a(\tau)=b(\tau),\quad
\dot b(\tau)+\si_1 b(\tau)+a(\tau)=
j(\tau),\qquad
 \dot c_n(\tau)=-i\Om_n(\tau) 
  c_n(\tau),
 \quad n\in\ov N,
  \ee
  where we denote 
  \be\la{ccOm}
c_n(\tau)=(c_{n,1}(\tau),c_{n,2}(\tau)),\qquad
 \Om_n(\tau)=
\left(\ba{cc} 0&\om_n(\tau)\\\ov\om_n(\tau)&0
\ea\right),\quad \om_n(\tau)=[\beta_n b(\tau)+\ga_n\cos\tau] e^{-i \tau}.
\ee
The system (\ref{MB31c}) can be written as
\be\la{XF}
\dot X(\tau)=F(X(\tau),\tau),\qquad 
X(\tau):=(a(\tau),b(\tau), c(\tau)),\qquad c(\tau):=(c_n(\tau):n\in\ov N).
\ee
This system (\ref{XF})  is   $2\pi$-periodic, i.e.,
\be\la{XFp}
F(X,\tau+2\pi)\equiv F(X,\tau).
\ee
The corresponding 
   Poincar\'e map  is defined by
  \be\la{Po2}
  P:X(0)\mapsto X(2\pi).
  \ee

 The following definition is relevant for the system  (\ref{XF}) 
  in the case of
sufficiently small pumping. 

\bd\la{dph}
{\rm
The parametric resonance occurs for 
the ground state (\ref{stsol}) when the condition
 (\ref{ph})  holds with $DP(X^0)$ instead of $D\cP(Y^0)$.
 }
 \ed

  We also need to check the additional condition for all the corresponding 
eigenvectors $\bX:=(\bba,\bb,(\bbc_n:n\in\ov N))$:
\be\la{cph}
(\bba,\bb)\ne 0,
\ee
which guarantees an exponential increment
of the Maxwell field. We have verified
this condition in the case of  ruby laser; see Remark \ref{rM}.

  \setcounter{equation}{0}
\section{Successive approximations}

We will construct the Poincar\'e map (\ref{Po2}) calculating successive approximations for 
the  equations  (\ref{MB31c}) on the interval $[0,2\pi]$
with  initial data
\be\la{iniw}
a(0),\quad b(0);\qquad c_n(0),\quad n\in\ov N,
\ee
which are close to the set of the ground states (\ref{zsol}), i.e., 
\be\la{iniw2}
|a(0)|+ |b(0)|+\sum_n ||c_{n,1}(0)|-1|+|c_{n,2}(0)|\le \ve.
\ee
Here $\ve>0$ is sufficiently small 
since we 
must to study instability of solutions which are close to these ground states.
The successive approximations are defined as follows.
\smallskip\\
{\bf The Schr\"odinger amplitudes.}
We define the approximations $c_n^{(k)}$ with $k=1,\dots$ as solutions of  equations
\be\la{MB31cac}
 \dot c_n^{(k)}(\tau)=-i\Om_n^{(k-1)}(\tau) 
  c_n^{(k)}(\tau),
 \quad n\in\ov N,
  \ee
  with the initial data (\ref{iniw}),
\be\la{iniMBc}
   c_n^{(k)}(0)=c_n(0),
 \quad n\in\ov N.
  \ee
Here we denote 
  \be\la{ccOmac}
  \Om_n^{(k-1)}(\tau)\!=\!
\left(\ba{cc} 0&\om_n^{(k-1)}(\tau)\\\ov\om_n^{(k-1)}(\tau)&0
\ea\right),
\qquad
\om_n^{(k-1)}(\tau)\!=\![\beta_n b^{(k-1)}(\tau)\!+\!\ga_n\cos\tau] e^{-i \tau}.
\ee 
To solve the  equations (\ref{MB31cac}), 
we replace them by the corresponding 
averaged equations
\be\la{avB}
\dot c_n^{(k)}(\tau)=-i\ti\Om_n^{(k-1)}
  c_n^{(k)}(\tau),
 \quad n\in\ov N,
\ee
where
\be\la{avOm}
\ti\Om_n^{(k-1)}:=\fr1{2\pi}\int_0^{2\pi} \Om_n^{(k-1)}(\tau)d\tau.
\ee
The matrix functions  $\Om_n^{(k-1)}(\tau)$ are sufficiently small by (\ref{abg}), so the
replacement is justified by Lemma \ref{lA}.
All the matrices (\ref{ccOmac}) and (\ref{avOm}) are skewsymmetric, and hence,
\be\la{UUU}
c_n^{(k)}(\tau)=U_n^{(k)}(\tau) c_n(0),
\ee
where $U_n^{(k)}(\tau)$ are unitary matrices.

\noindent
{\bf The Maxwell amplitude.} We define the approximations $a^{(k)}(\tau)$ with $k=1,\dots$ as solutions of the equations
\be\la{MB31ca}
\ddot a^{(k)}(\tau)+\si_1 \dot a^{(k)}(\tau)+a^{(k)}(\tau)=
j^{(k)}(\tau),\qquad j^{(k)}(\tau)\!:=\!\sum_n\al_n
\rIm\Big\{\ov c_{n,1}^{(k)}(\tau)c_{n,2}^{(k)}(\tau)  e^{\!-\!i\tau}\Big\},
  \ee
  with the initial data (\ref{iniw}):
  \be\la{iniMB}
  a^{(k)}(0)=a(0),\qquad \dot a^{(k)}(0)=b(0).
    \ee
    Let us denote  $\vka:=\si_1/2\approx 10^{-7}$.
The solution of
the Maxwell equation (\ref{MB31ca}) 
 is given by  the convolution
\be\la{MB31in}
a^{(k)}(\tau)=
a_0(\tau)+
\int_0^\tau j^{(k)}(\tau')E(\tau-\tau')d\tau',\qquad 
a_0(\tau):=a(0)\dot E(\tau)+[\dot a(0)+2\vka a(0)]E(\tau),
\ee
where $E(\tau)$ is
the retarded fundamental solution,
 \be\la{funs}
 \ddot E(\tau)+2\vka\dot E(\tau)+E(\tau)=\de(\tau),\quad\tau\in\R;\qquad E(\tau)=0,
 \quad\tau<0.
 \ee
This fundamental solution is expressed as follows.
Denote by $\lam_\pm$ the roots of characteristic equation 
\be\la{chaeq}
\lam^2+2\vka\lam+1=0
\ee
which are 
\be\la{lam}
\lam_\pm= -\vka\pm \sqrt{\vka^2-1}=
-\vka\pm i+
O(\vka^2).
\ee
Thus, $\rRe\lam_\pm=-\vka<0$, and 
neglecting terms which are $O(\vka^2)$, we obtain that
the functions
\be\la{Msol}
a_1(\tau)=\fr{\lam_+e^{\lam_-\tau}-\lam_-e^{\lam_+\tau}}{\lam_+-\lam_-}=
e^{-\vka\tau}\cos\tau+O(\vka^2),
\qquad
a_2(\tau)=\fr{e^{\lam_+\tau}-e^{\lam_-\tau}}{\lam_+-\lam_-}=
e^{-\vka\tau}\sin\tau+O(\vka^2),
\ee
are solutions of the homogeneous Maxwell equation (\ref{MB31ca}) with initial data
$(1,0)$ and $(0,1)$ respectively.
Finally, 
the fundamental solution is the real-valued function
\be\la{funs0}
E(\tau)=\theta(\tau)a_2(\tau)=
\theta(\tau)e^{-\vka\tau}\sin\tau+O(\vka^2),\qquad \tau\in\R.
\ee

\subsection{The first-order approximation}

{\bf The Schr\"odinger amplitudes.} Formulas (\ref{ccOmac})-- (\ref{avOm}) with $k=1$ give the averaged matrices
\be\la{zeOmjc}
\ti\Om^{(0)}_n(\tau)=
\left(\ba{cc} 0&-i\fr{\ga_n}2
\\-i\fr{\ga_n}2&0
\ea\right).
\ee
Hence, 
equation (\ref{MB31cac}) 
with $k=1$
reads as the system in
(\ref{MB31})  with $ \fr{\ga_n}2$ instead of $\om_n(\tau)$:
\be\la{MB32a}
\left\{\ba{rcl}
 \dot c_{n,1}^{(1)}(\tau)&=&-i\fr{\ga_n}2
  c_{n,2}^{(1)}(\tau)
\\
\\
\dot c_{n,2}^{(1)}(\tau)&=&- i\fr{\ga_n}2
  c_{n,1}^{(1)}(\tau)
 \ea\right|, \qquad n\in\ov N.
\ee
Its solutions are
\be\la{1apr}
c_{n,1}^{(1)}(\tau)=d_{n,1}^{(1)}\cos \fr{\ga_n\tau}2+d_{n,2}^{(1)}\sin\fr{\ga_n\tau}2,\qquad
c_{n,2}^{(1)}(\tau)=
-id_{n,1}^{(1)}\sin\fr{\ga_n\tau}2+id_{n,2}^{(1)}\cos\fr{\ga_n\tau}2,
\ee
where 
\be\la{dcdc}
d_{n,1}^{(1)}=c_{n,1}(0),\qquad d_{n,2}^{(1)}=-ic_{n,2}(0).
\ee
 Hence, the matrix  in the formula 
 (\ref{UUU})  with $n=1$ reads 
\be\la{PC}
U_n^{(1)}(\tau)=\left(\ba{cc}\cos\fr{ \ga_n\tau}2& -i\sin\fr{ \ga_n\tau}2
\\\\
-i\sin\fr{\ga_n\tau}2&\cos \fr{\ga_n\tau}2
\ea
\right)+O(\ga_n^2)
\ee
by Lemma \ref{lA}.

\noindent
{\bf The current.}
Substituting (\ref{1apr})  
and (\ref{dcdc})
in the second formula of
 (\ref{MB31ca}) with $k=1$,  we obtain
 \beqn\la{jn2c}
\!\!\!
j^{(1)}(\tau)\!\!&\!\!=\!\!&\!\!\sum_n\al_n
\rIm\Big\{\ov c_{n,1}^{(1)}(\tau)c_{n,2}^{(1)}(\tau)  e^{\!-\!i\tau}\Big\}
 \nonumber\\
   \nonumber\\
\!\!&\!\!=\!\!&\!\!\sum_n\al_n
\rIm\Big\{[\ov d_{n,1}^{(1)}\cos \fr{\ga_n\tau}2+\ov d_{n,2}^{(1)}\sin\fr{\ga_n\tau}2]
[-id_{n,1}^{(1)}\sin\fr{\ga_n\tau}2+id_{n,2}^{(1)}\cos\fr{\ga_n\tau}2]
  e^{-i\tau}\Big\}
  \nonumber\\
   \nonumber\\
   \!\!&\!\!=\!\! &\!\!
   \sum_n\al_n
\rIm\Big\{[i\ov d_{n,1}^{(1)} d_{n,2}^{(1)}-i|d_{n,1}^{(1)}|^2\sin\fr{\ga_n\tau}2+i|d_{n,2}^{(1)}|^2\sin\fr{\ga_n\tau}2]
  e^{-i\tau}\Big\}+O(\sum_n\al_n\ga_n^2)
   \nonumber\\
   \nonumber\\
 \!\!  &\!\!=\!\!&\!\!
  \sum_n\al_n
\rRe\Big\{[\ov d_{n,1}^{(1)} d_{n,2}^{(1)}-|d_{n,1}^{(1)}|^2\sin\fr{\ga_n\tau}2+|d_{n,2}^{(1)}|^2\sin\fr{\ga_n\tau}2]
  e^{-i\tau}\Big\}+O(\sum_n\al_n\ga_n^2)
  \nonumber\\
   \nonumber\\
  \!\! &\!\!=\!\!&\!\!
   \sum_n\al_n
\Big[\rIm\Big\{\ov c_{n,1}(0) c_{n,2}(0)e^{-i\tau}\Big\}
+[|c_{n,2}(0)|^2-|c_{n,1}(0)|^2]\fr{\ga_n\tau}2
  \cos\tau\Big]\!\!+\!O(10^{-17}),\,\tau\!\in\![0,2\pi],\qquad
 \eeqn  
 since
 $\sum_n\al_n\ga_n^2\sim 10^{-17}$ by (\ref{abg}).
\smallskip

\noindent{\bf Notation}. 
For each number $\ve>0$, the equality
$r=O(\ve)$ 
means that $|r|\le 10 \ve$.
 \smallskip
 
 \noindent
{\bf The Maxwell amplitude.}
Now (\ref{MB31in}) with $k=1$  becomes
\be\la{MB31in1}
a^{(1)}(\tau)=
 a_0(\tau) +
\int_0^\tau j^{(1)}(\tau')E(\tau-\tau')d\tau'.
\ee
Substituting here the current  (\ref{jn2c}), we obtain
\be\la{MB31in2}
a^{(1)}(\tau)\!=\!
 a_0(\tau)\!+\! \sum_n\al_n
\Big[\rIm\Big\{\ov c_{n,1}(0) c_{n,2}(0)I_1(\tau)\Big\}
+[|c_{n,2}(0)|^2-|c_{n,1}(0)|^2] \ga_n I_2(\tau)\Big]\!+\!
O(10^{-17}),
\ee
where we denote 
\beqn\la{I12}
I_1(\tau)&:=&\int_0^\tau e^{-i\tau'}E(\tau-\tau')d\tau'
=
-\fr i2\sin\tau-
\fr\tau{2i}e^{-i\tau}+\vka\Big[\fr{-\sin\tau+\tau e^{i\tau}}{4}+\fr{\tau^2e^{-i\tau}}{4i}\Big]+O_2.
\nonumber\\
\nonumber\\\nonumber
I_2(\tau)&:=&\int_0^\tau 
\fr{ \tau'}2  \cos\tau' E(\tau-\tau')d\tau'=
\fr{1}4 
  \Big[\fr{\tau^2}2\sin\tau
  +
  \fr\tau2
  \cos\tau
    -\fr1{2}
    \sin\tau
  \Big]+O_2,
\eeqn
where $O_2:=O(\vka^2)=O(10^{-14})$.
The details of calculation can be found 
 in Appendix \ref{aB}.

\subsection{The second-order approximation}
{\bf The Schr\"odinger amplitudes.}
Formulas    
 (\ref{ccOmac}) with $k=2$ give 
\be\la{Omom2}
\Om_n^{(1)}(\tau)=
\left(\ba{cc} 0&\om_n^{(1)}(\tau)\\\ov\om_n^{(1)}(\tau)&0
\ea\right),\qquad 
 \om_n^{(1)}(\tau)=[\beta_n\dot a^{(1)}(\tau)+\ga_n\cos\tau] e^{-i \tau}.
\ee
Let us calculate the averaged matrices $\ti\Om_n^{(1)}$.
First,   (\ref{MB31in2}) and (\ref{MB31in}) imply that
\be\la{MB31in3}
\nu:=\fr1{2\pi}
\int_0^{2\pi} \dot a^{(1)}(\tau)e^{-i \tau}d\tau=\nu_1+\nu_2,
\ee
where 
\beqn
\nu_1&:=&\fr1{2\pi}
\int_0^{2\pi} \dot a_0(\tau)e^{-i \tau}d\tau=\fr1{2\pi}\int_0^{2\pi}
\{
a(0)\ddot E(\tau)+[\dot a(0)+2\vka a(0)]\dot E(\tau)\}e^{-i \tau}d\tau,
\nonumber\\
\nonumber\\\nonumber
\nu_2&:=&
\fr1{2\pi}
\int_0^{2\pi} 
  \sum_n\al_n
\Big[\rIm\Big\{\ov c_{n,1}(0) c_{n,2}(0)
I_1'(\tau)e^{-i\tau}
\Big\}
+[|c_{n,2}(0)|^2-|c_{n,1}(0)|^2] \ga_n
 I_2'(\tau)e^{-i\tau}
\Big]d\tau
\eeqn
First, let us calculate $\nu_1$.
Neglecting errors $O(\vka^2)=O(10^{-14})$, we obtain:
\beqn\la{aver3}
\int_0^{2\pi}\dot E(\tau)e^{-i \tau}d\tau
\!\!\!\!&\!\!\!\!\approx\!\!\!\!&\!\!\!\!i\int_0^{2\pi}E(\tau)e^{-i \tau}d\tau
\approx i\int_0^{2\pi} e^{-\vka\tau}\sin\tau\,e^{-i \tau}d\tau=\fr12\int_0^{2\pi} e^{-\vka\tau}(e^{i \tau}-e^{-i \tau})e^{-i \tau}d\tau
\nonumber\\
\nonumber\\
\!\!\!\!&\!\!\!\!=\!\!\!\!&\!\!\!\!
-[\fr1{2\vka}+\fr1{2\vka+4i}](e^{-2\vka\pi}-1)
=
-\fr{2\vka+2i}{\vka+2i}\,\,\fr{-\vka\pi+(\vka\pi)^2+O(\vka^3)}{\vka}
\nonumber\\
\nonumber\\
&\approx&
-[1+\fr{\vka}{\vka+2i}](-\pi+ \vka\pi^2)
\approx
-[1+\fr{\vka}{2i}](-\pi+\vka\pi^2)\approx
\pi+\vka(\fr\pi{2i}-\pi^2).\qquad\qquad\qquad
\eeqn
Similarly, using (\ref{funs}), (\ref{funs0}), and  (\ref{aver3}), we obtain
\beqn\la{aver4}
\int_0^{2\pi}\ddot E(\tau)e^{-i \tau}d\tau
&=&\int_0^{2\pi}[-2\vka\dot E(\tau)-E(\tau)]e^{-i \tau}d\tau
\approx-(i2\vka+1)\int_0^{2\pi}E(\tau)e^{-i \tau}d\tau
\nonumber\\
\nonumber\\
&\approx&i(i2\vka+1)[\pi+\vka(\fr\pi{2i}-\pi^2)]
=
(-2\vka+i)[\pi+\vka(\fr\pi{2i}-\pi^2]
\nonumber\\
\nonumber\\
&\approx&
i\pi+\vka[-2\pi+\fr\pi{2}-i{\pi^2}]
=
i\pi-\vka[\fr{3\pi}2+i\pi^2].
\eeqn
As a result,
\be\la{aver22}
\left.\ba{rcl}
\nu_1&\approx&
a(0)\Big[ \fr i2-\vka[\fr{3}4+i\fr{\pi}2]\Big]+[\dot a(0)+2\vka a(0)]
\Big[  \fr12+\vka(\fr1{4i}-\fr\pi2) \Big]
\\\\
&\approx&
a(0)\Big[\fr i2+\vka (\fr14-i\fr{\pi}2)\Big]
+\dot a(0)
\Big[  \fr12+\vka(\fr1{4i}-\fr\pi2) \Big]
=\nu_{11}+i\nu_{12},
\\\\
\nu_{11}& =&\fr\vka 4a(0)+\fr12(1-\vka\pi)\dot a(0),\qquad
\nu_{12}=\fr12(1-\vka\pi)a(0)-\fr\vka4\dot a(0)
\ea\right|.
\ee

It remains to note that
\be\la{J10}
\nu_2:=
\sum_n\al_n\Big[\rIm\Big\{\ov c_{n,1}(0) c_{n,2}(0) J_1
\Big\}+[|c_{n,2}(0)|^2-|c_{n,1}(0)|^2] \ga_n J_2\Big],
\ee
where $J_1$ and $J_2$ are the integrals
\be\la{JI12}
J_1:=\ds\fr1{2\pi}\int_0^{2\pi}  I_1'(\tau)e^{-i\tau}d\tau\approx0,
\qquad
J_2:=\ds
\ds\fr1{2\pi}\int_0^{2\pi}  I_2'(\tau)e^{-i\tau}d\tau\approx\fr{\pi^2}{12}.
\ee
We calculate these integrals in Appendix \ref{aB}.
As a result, the averaged matrix reads
\be\la{aver23}
\ti\Om_n^{(1)}(\nu)=\left(\ba{cc} 0&\ti\om_n^{(1)}\\\ov{\ti\om}_n^{(1)}&0
\ea\right),\qquad 
 \ti\om_n^{(1)}=\beta_n\nu +\fr{\ga_n}2.
\ee
Now equation (\ref{avB}) with $k=2$ reads as the system
\be\la{MB322}
\left\{\ba{rcl}
 \dot c_{n,1}^{(2)}(\tau)&=&-i\ti\om_n^{(1)}
  c_{n,2}^{(2)}(\tau)
\\
\\
\dot c_{n,2}^{(2)}(\tau)&=&- i\ov{\ti\om}_n^{(1)} 
  c_{n,1}^{(2)}(\tau)
 \ea\right|, \qquad n\in\ov N.
 \ee
Its solution reads similarly to (\ref{1apr}):
\be\la{1apr2}
c_{n,1}^{(2)}(\tau)\!=\!d_{n,1}^{(2)}\cos |\ti\om_n^{(1)}|\tau\!+\!d_{n,2}^{(2)}\sin|\ti\om_n^{(1)}|\tau,\quad
c_{n,2}^{(2)}(\tau)\!=\!
-\!id_{n,1}^{(2)}
\fr{|\ti\om_n^{(1)}|}{\ti\om_n^{(1)}}
\sin|\ti\om_n^{(1)}|\tau\!+\!id_{n,2}^{(2)}\fr{|\ti\om_n^{(1)}|}{\ti\om_n^{(1)}}\cos|\ti\om_n^{(1)}|\tau,
\ee
where 
\be\la{dcdc2}
d_{n,1}^{(2)}=c_{n,1}(0),\qquad d_{n,2}^{(2)}=-i\fr{\ti\om_n^{(1)}}{|\ti\om_n^{(1)}|}c_{n,2}(0).
\ee
Thus, (\ref{1apr2}) can be rewritten as
\be\la{1apr2r}
\left\{\ba{rcl}
c_{n,1}^{(2)}(\tau)&=&c_{n,1}(0)\cos |\ti\om_n^{(1)}|\tau-is_nc_{n,2}(0)\sin|\ti\om_n^{(1)}|\tau
\\\\
c_{n,2}^{(2)}(\tau)&=&\!\!\!\!\!\!
-i\ov s_n c_{n,1}(0)
\sin|\ti\om_n^{(1)}|\tau+c_{n,2}(0)
\cos|\ti\om_n^{(1)}|\tau,\
\ea\right|,
\ee
where $s_n=\fr{\ti\om_n^{(1)}}{|\ti\om_n^{(1)}|}$.
Hence, the matrix in the formula  (\ref{UUU}) with $k=2$ reads
\be\la{Uk2}
U_n^{(2)}(\tau)=\left(\ba{cc}\cos |\ti\om_n^{(1)}|\tau&-is_n\sin |\ti\om_n^{(1)}|\tau
\\\\
-i\ov s_n\sin|\ti\om_n^{(1)}|\tau&\cos |\ti\om_n^{(1)}|\tau
\ea
\right).
\ee
 
\noindent{\bf The current.} 
Expanding (\ref{1apr2r}) into the Taylor series, we obtain
\be\la{1apr2rT}
\left\{\ba{rcl}
c_{n,1}^{(2)}(\tau)&=&c_{n,1}(0)-ic_{n,2}(0)\ti\om_n^{(1)}\tau+O(|\ti\om_n^{(1)}|^2)
\\\\
c_{n,2}^{(2)}(\tau)&=&\!\!\!\!\!\!
-\!i c_{n,1}(0)
\ov{\ti\om}_n^{(1)}\tau+c_{n,2}(0)
+O(|\ti\om_n^{(1)}|^2)
\ea\right|,
\ee
Here the error    $O(|\ti\om_n^{(1)}|^2)\sim 10^{-14}$  uniformly in $\tau\in[0,2\pi]$.
    Indeed, $ \ti\om_n^{(1)}=\beta_n\nu +\fr{\ga_n}2$, where 
 $\nu$ is expressed by (\ref{MB31in3}) and (\ref{aver22})--(\ref{JI12}): 
\be\la{nu}
\left\{\ba{rcl}
 \nu&=&\nu_{11}+i\nu_{12}+\nu_2\\\\
 \nu_{11} &\approx&\ds \fr{b(0)}2+\fr\vka4[a(0)-2\pi b(0)],\qquad
\nu_{12} \approx
\fr {a(0)}2-\fr\vka4[2\pi a(0)+b(0)]
\\\\
\nu_2&\approx&J_2
 \sum_n\al_n\ga_n [|c_{n,2}(0)|^2-|c_{n,1}(0)|^2] 
\ea\right|,
\ee
Hence at the points (\ref{zsol}), we have $\nu_{11}=\nu_{12}=0$, so
\be\la{nunu}
\beta_n\nu=\beta_n\nu_2=-\beta_nJ_2\sum_n\al_n\ga_n\sim 10^{-12}
\ee
by
 (\ref{abg}) and (\ref{JI12}). Therefore, 
 \be\la{om1nu}
  \ti\om_n^{(1)}=\beta_n\nu +\fr{\ga_n}2\approx \fr{\ga_n}2\sim 10^{-7}.
  \ee
 Finally, substituting (\ref{1apr2rT})
  into  the second formula of
 (\ref{MB31ca}) with $k=2$,  we obtain similarly to (\ref{jn2c}), 
   \beqn\la{jn2c3}
j^{(2)}(\tau)&=&\sum_n\al_n
\rIm\Big\{\ov c_{n,1}^{(2)}(\tau)c_{n,2}^{(2)}(\tau)  e^{-i\tau}\Big\}
\nonumber\\
      \nonumber\\
 &=&
  \sum_n\al_n
\rIm\Big\{\ov c_{n,1}(0) c_{n,2}(0)e^{-i\tau}
+i\ov{\ti\om}_n^{(1)}[|c_{n,2}(0)|^2 -|c_{n,1}(0)|^2]\tau
   e^{-i\tau}  
\Big\}+O(|\ti\om_n^{(1)}|^2)
    \eeqn

 \noindent{\bf The Maxwell amplitude.} 
The last formula of (\ref{MB31in}) with $k=2$ gives
  \be\la{Pmap2t}
 \left\{\ba{rcl}
 a^{(2)}(\tau)&=&a_0(\tau)+
\ds \int_0^{\tau} j^{(2)}(\tau') E(2\pi-\tau')d\tau'
 \\\\
b^{(2)}(\tau)&=&\dot a_0(\tau)+
\ds \int_0^{\tau} j^{(2)}(\tau') \dot E(2\pi-\tau')d\tau'
 \ea\right|.
 \ee

 \setcounter{equation}{0}
 \section{The Poincar\'e map}
 Let us denote 
 \be\la{abc}
 a^0:=a(0),\,\,\, b^0:=\dot a(0),\,\,\, c_n^0:=c_n(0);\qquad
 a:=a^{(2)}(2\pi),\,\, b:=\dot a^{(2)}(2\pi),
 \,\,\, c_n:=c_n^{(2)}(2\pi).
 \ee
 Then  (\ref{Pmap2t}) gives
 \be\la{Pmap2}
 a=a_0(2\pi)+
\ds \int_0^{2\pi} j^{(2)}(\tau') E(2\pi-\tau')d\tau',
\qquad
b=\dot a_0(2\pi)+
\ds \int_0^{2\pi} j^{(2)}(\tau') \dot E(2\pi-\tau')d\tau'.
  \ee
 Let us calculate these expressions. First, (\ref{MB31in}) and (\ref{funs}), (\ref{funs0}) imply that
 \beqn\la{dota}
 a_0(2\pi)=a^0\dot E(2\pi),\quad 
 \dot a_0(2\pi)&=&a^0\ddot E(2\pi)+[b^0+\si_1 a^0]\dot E(2\pi)
 \nonumber\\
 \nonumber\\
 &=&
 -a^0 E(2\pi)
 +b^0\dot E(2\pi)\approx
 b^0\dot E(2\pi)+O_2,
 \eeqn
 where $O_2:=O(\vka^2)\sim 10^{-14}$.
 Second, substituting $\ti\om_n^{(1)}=\beta_n\nu +\fr{\ga_n}2$ into
 (\ref{jn2c3}), we obtain
  \beqn\la{jwabc}
  j^{(2)}(\tau)&\approx&  \sum_n\al_n
\rIm\Big\{\ov c_{n,1}^0 c_{n,2}^0e^{-i\tau}\Big\}
\nonumber\\
  \nonumber\\
  &+&
   \sum_n\al_n\beta_n
   [|c_{n,2}^0|^2 -|c_{n,1}^0|^2]
   \rRe\big\{
 \ov \nu\tau
   e^{-i\tau}\big\}
+
\fr12 \sum_n\al_n \ga_n
[|c_{n,2}^0|^2-|c_{n,1}^0|^2] 
\tau\cos\tau.
  \eeqn
Introducing this current  into (\ref{Pmap2}),
 we obtain the formulas for $a$ and $b$ expressed via the following integrals
 which are calculated in Appendix \ref{aB}:
 \be\la{AB}
 ~\!\!\!
\!\!\!\!\!\!\!\!\!\left\{\!\!\!\ba{ll}
  A_1\!:=\!\ds\!\int_0^{2\pi}\!\!\!\! e^{-i\tau'} E(2\pi\!-\!\tau')d\tau'\!,\,A_2\!:=\!\!\int_0^{2\pi}\!\!\!\! \tau' e^{-i\tau'} E(2\pi\!-\!\tau')d\tau'\!,\,A_3\!:=\!\!\int_0^{2\pi} \!\!\!\!\tau'\cos\tau' E(2\pi\!-\!\tau')d\tau'\!=\!\rRe A_2
  \\\\
   B_1\!=\!\ds\!\int_0^{2\pi}\!\!\!\! e^{-i\tau'} \dot E(2\pi\!-\!\tau')d\tau'\!,\,B_2\!:=\!\!\int_0^{2\pi}\!\!\!\! \tau' e^{-i\tau'}\dot E(2\pi\!-\!\tau')d\tau'\!,\,B_3\!:=\!\!\int_0^{2\pi}\!\!\!\! \tau' \cos\tau'\dot E(2\pi\!-\!\tau')d\tau'\!=\!-\rIm B_2
   \ea\!\!\!\right|.
 \ee
With this notation,  (\ref{Pmap2}) becomes
\be\la{Pmap4}
 \left\{\ba{rcl}
 a&\approx&a_0(2\pi)+
 \sum_n\al_n
\rIm\Big\{\ov c_{n,1}^0 c_{n,2}^0A_1\Big\}
\\\\
  &&+
   \sum_n\al_n\beta_n
   [|c_{n,2}^0|^2 -|c_{n,1}^0|^2]\rRe\big\{
\ov \nu A_2\big\}
+
\fr12 \sum_n\al_n \ga_n
[|c_{n,2}^0|^2-|c_{n,1}^0|^2] 
A_3.
 \\\\
b&\approx&\dot a_0(2\pi)+
 \sum_n\al_n
\rIm\Big\{\ov c_{n,1}^0 c_{n,2}^0B_1\Big\}
\\\\
  &&+
   \sum_n\al_n\beta_n[|c_{n,2}^0|^2 -|c_{n,1}^0|^2]\rRe\big\{
 \ov \nu B_2\big\}
+
\fr12 \sum_n\al_n \ga_n
[|c_{n,2}^0|^2-|c_{n,1}^0|^2] 
B_3.
 \ea\right|.
 \ee
  
\br
{\rm
i) The Poincar\'e map $P$ is given by formulas (\ref{Pmap4}) and (\ref{1apr2rT}).
\smallskip\\
ii)
The second-order approximations (\ref{Pmap4}) are obtained  solving
  the equation  
 (\ref{MB31ca}) with $k=2$,  initial data (\ref{iniw}), and the current
 (\ref{jwabc}).
 At the ``ground states" (\ref{zsol}),  
we have $|c_{n,1}^0|^2=1$ and 
$|c_{n,2}^0|^2=0$, so
the formula (\ref{jwabc})
takes the form
 \beqn\la{jwabc2}
  j^{(2)}(\tau)= -   \sum_n\al_n\beta_n
\tau\,\rIm\Big\{ \ov\nu 
   e^{-i\tau}\Big\}
-
\fr12 \sum_n\al_n \ga_n 
\tau\cos\tau.
  \eeqn

}
\er

\setcounter{equation}{0}
\section{Synchronisation of molecular currents}
The key peculiarity of the perturbative  calculations is
the appearance of the summation
\be\la{SSig}
\bS=\sum_n\al_n\beta_n
\ee
in the second-order approximation
of the 
 total molecular current (\ref{jwabc2}).
This appearance is due to the Taylor approximation (\ref{1apr2rT}).
 Also note that the same summation reappears in
  (\ref{Su}).
It is of crucial importance that all terms of the sum  $\bS$ {\it are nonnegative}
 since
\be\la{abnn}
\al_n\beta_n= \fr{2 c}{\Om_p}\bP_n  \bX(x_n)
 \ds\fr{\bP_n\bX(x_n)}{\hbar c}\ge 0,
\ee
 and
the sum is a non-negligible positive number, see  (\ref{probs2}) and (\ref{probs22cr}) below. 
This means the {\it synchronisation}  of the main parts of all molecular currents 
in the total current (\ref{jwabc}), (\ref{jwabc2})
 {\it independently} of orientations of the polarization $\bP_n$ and 
of the Maxwell mode $\bX(x_n)$  at the location of the molecules. 

\br\rm
This {\it synchronisation of molecular currents} is provisionally in line with the {\it superradiance} and with
 the role of the {\it stimulated emission} in the laser action.
 Note that  the conjecture on appropriate kind of synchronisation
  in the laser action  has been put forward repeatedly, see, e.g., \ci{PRK2001}.
  \er
Note that we must obtain {\it the value of the summation} $\bS$ for the calculation of the 
Poincar\'e map, although the summands are  random values.
Indeed, in  (\ref{albege}),
all the parameters $\bP_n$, $\bX(x_n)$, $\bba_p(x_n)$
  with $n\in\ov N$
 can be considered as  {\it independent random values}, and
 \be\la{E0}
\cE\bX(x_n)=\cE\bba_p(x_n)=0.
\ee
We will calculate the summation $\bS$ in both cases of polycrystalline and monocrystalline medium  by probabilistic arguments as follows.
We will consider two cases of amorphous and crystallin medium separately.

\noindent{\bf I. Amorphous medium.} By our assumption,
 $|\bP_n|=|\bP|$ does not depend on $n$, so 
\be\la{Pe}
(\bP_n\cdot e_1)^2+(\bP_n\cdot e_1)^2+(\bP_n\cdot e_1)^2=|\bP|^2.
\ee
Moreover, 
the dipole momenta $\bP_n$
are distributed uniformly over the angles, and hence,
\be\la{Pe2}
\cE(\bP_n\cdot e_1)^2=\fr13|\bP|^2.
\ee
By Remark \ref{rH}, we can apply 
the {\it law of  large numbers}
for weakly dependent random values \ci{IL1971}
 (recall that $N\sim 10^{20}$), and obtain that
 $\bS\approx N\cE \al_n\beta_n$.
Moreover,
 $\bP_n$ and $\bX(x_n)$ are independent.
 Hence, using (\ref{dipap}), 
  we obtain that
\beqn\la{probs2}
 \bS\!\!&\!\!\approx\!\!&\!\!N\cE  \al_n\beta_n=N\fr2{\Om_p\hbar}  \cE (\bP_n\bX(x_n))^2
=N\fr2{\Om_p\hbar}  \fr{\cE|\bP_n|^2}3 \cE\bX^2(x_n)
\nonumber\\
\nonumber\\
\nonumber\\
\!\!&\!\!=\!\!&\!\!
\fr2{\Om_p\hbar}\fr{|\bP|^2}3  \sum_n \bX^2(x_n)
\approx \fr{2|\bP|^2}{3\Om_p\hbar} \fr {N}{|V_a|}\int_{V_a}  \bX^2(x)dx=
 \fr{2|\bP|^2}{3\Om_p\hbar}d\fr {|V_a|}{|V|},\quad d:= \fr {N}{|V_a|}
\eeqn
by the normalization of $\bX$.
Here the value $|V_a|/|V|$ for the integral follows from 
  Shnirelman's 
{\it Quantum Ergodic theorem} \ci{S2011,S1974,S1993}.
Note that 
this value obviously holds for the eigenmodes (\ref{eigcubo}) with sufficiently large eigenvalues.
For the case of ruby laser (\ref{abg})
we obtain $d\approx 3\times 10^{19}$ and 
\be\la{probs22}
\bS
\sim 10^{-5},
\ee
which agrees with 
estimate $\bS\sim 10^{20}\times10^{-25}$.
This agreement confirms the obtained formula (\ref{probs2}).
Moreover, $\bS$ is a Gaussian random value since
$\al_n\beta_n$ are independent and identically distributed.
\smallskip\\
{\bf II. Crystallin medium.}  
In this case
 the vectors $\bP_n=\bP$  do not depend on $n$.
We assume additionally that the mode $\bX(x)$  
is linearly polarised.
Now (\ref{probs2}) changes to
\be\la{probs22cr}
\bS\approx
\fr2{\Om_p\hbar}|\bP|^2\cos^2\vp  \sum_n \bX^2(x_n)
\approx 
 \fr{2|\bP|^2\cos^2\vp}{\Om_p\hbar}d\fr {|V_a|}{|V|}\approx 32\fr {|V_a|}{|V|}10^{-5}\cos^2\vp
 \sim 10^{-5},
\ee
where $\vp$ is the angle between $\bP$ and $\bX(x)$.

 \setcounter{equation}{0}
\section{Factordynamics and the Hopf fibration}\la{sHR}
The system (\ref{MB31c}) is invariant with respect to the 
 {\it symmetry gauge group} $G=[U(1)]^N$
acting
on the phase space $\cX=\R^2\oplus \C^{2N}$
by the formula similar to (\ref{actG}):
\be\la{actG2}
(e^{i\Theta_1},\dots, e^{i\Theta_{N}})
(a, b,(c_{11},c_{12}),\dots,  (c_{N,1},c_{N,2}))=
(a, b, e^{i\Theta_1}(c_{11},c_{12}),\dots,  
e^{i\Theta_{N}}(c_{N,1},c_{N,2})).
\ee
This action commutes with the dynamics (\ref{MB31c}),
hence the latter induces the corresponding dynamics   on the factorspace 
$\cX_*=\cX/G$. This factorspace is the base of the fibration
$\cX\to \cX_*$.
The action  (\ref{actG2})
does not change the Maxwell component $(a,b)$. 
Moreover, this action commutes with 
the Poincar\'e map $P$. 
Hence, $P$ induces the map $P_*$ and the factorspace $\cX_*$.
Accordingly, the instability and the parametric resonance
(\ref{ph}) must be checked for the induced factordynamics only.
So, we must calculate 
  the differential
$DP_*(X^0_*)$, where 
 the point $X^0_*$ corresponds
to $X^0$ in this factorization (see (\ref{stsol})).

\noindent{\bf Local coordinates.}
To calculate  $DP_*(X^0_*)$,  we need  suitable local coordinates
 on the factorspace.
The charge conservation (\ref{Blcc}) means that 
\be\la{S3}
 (c_{n,1}(\tau),c_{n,2}(\tau))\in S^3,
\ee
where $S^3$ is the unit sphere in $\C^2$.
The sphere is the Hopf fibration with  fibers 
\be\la{Fib}
F=(c_1e^{i\Theta},c_2e^{i\Theta}),\qquad\Theta\in  [0,2\pi]
\ee
and the base  diffeomorphic to the unit sphere $S^2$.
Denote by $(c_1,c_2)_*\in S^2$ the Hopf projection of a point
$(c_1,c_2)\in S^3$.
We choose the following  invariant coordinate on the base:
\be\la{z0} 
z:=\ov c_1 c_2.
\ee
Using (\ref{Blcc}),
it is easy to check that 
\be\la{popul}
 |c_1|+ |c_2|=\sqrt{1+2|z|},\quad  |c_1|- |c_2|=\pm\sqrt{1-2|z|}.
 \ee
Hence,
\be\la{popul2}
2 |c_1|=\sqrt{1+2|z|}+\sqrt{1-2|z|},\quad  2|c_2|=\sqrt{1+2|z|}-\sqrt{1-2|z|}
 \ee
and
\be\la{popul3}
4 |c_1|^2=2+2\sqrt{1-4|z|^2},\quad  4|c_2|^2=2-2\sqrt{1-4|z|^2}.
 \ee
In particular, $ |c_1|^2$ and  $|c_2|^2$
are smooth functions on the disk $D:=|z| < \fr12$, and
\be\la{c20}
\na_z |c_l|^2|_{z=0}=0,
\ee
where $\na_z$ denotes the gradient with respect to $\rRe z$ and
 $\rIm z$.

The coordinate $z$ maps $S^3$  onto the closed disk $\ov D$,
and it is a local coordinate  on the region  of  $S^3$ outside the set  
$|c_1|^2=|c_2|^2=1/2$.
 However, 
 the inverse map $D\to S^2$ is two-valued.
 Indeed, denote by $(c_1,c_2)_*\in S^2$ the Hopf projection of a point
$(c_1,c_2)\in S^3$.
 One branch of the inverse map
  is a diffeomorphism of  $D$ onto the  neighborhood $S^2_+\subset S^2$ of the point
   $(1,0)_*$ (where $z=0$), and another 
branch  is a diffeomorphism of  $D$ onto the neighborhood $S^2_-\subset S^2$ of 
 the point $(0,1)_*$ (where also  $z=0$):
 \be\la{popul4}
\left\{\ba{rcl}
|c_1|>|c_2|&{\rm for}& (c_1,c_2)_*\in S^2_+\\\\
|c_1|<|c_2|&{\rm for}& (c_1,c_2)_*\in S^2_-
\ea\right|.
\ee
The base $S^2$ is the gluing of $S^2_\pm$ along their boundaries,
 where $|z|=\fr12$ and $|c_1|^2=|c_2|^2=\fr12$.
\smallskip

 Thus, the Hopf representation of the dynamics (\ref{MB31}) 
 uses  gauge-invariant  
 variables $z_n=\ov c_{n,1}c_{n,2}$. This
 reduces the number of variables twice that considerably simplifies the calculations.
In these new variables, the current (\ref{jbk}) 
and the population invertions $I_n:=|c_{n,2}|^2-|c_{n,1}|^2$
read as
\be\la{tajom2}
j(\tau)=\sum_n\al_n
\rIm\Big\{z_{n}(\tau)  e^{-i\tau}\Big\},\qquad I_n=-\sqrt{1-4|z_n|^2}
\ee
by  (\ref{c20}).
  So,
 the system (\ref{MB31}) becomes
 \be\la{zls}
 \dot a=b,\quad \dot b(\tau)+\si_1 b(\tau)+a(\tau)=
j(\tau),\qquad
\dot z_n
=
-i\ov\om_n\sqrt{1-4|z_n|^2}.
\ee

\setcounter{equation}{0}
\section{Differential of the Poincar\'e map on the factorspace} 
It remains to calculate the differential (\ref{Mder}).
We choose  coordinates (\ref{z0}) on the bases
of the Hopf fibration
 for each active molecule, and denote
\be\la{proco}
z_n^0:=\ov c_{n,1}^0 c_{n,2}^0=\ov c_{n,1}(0) c_{n,2}(0),\qquad
z_n:=\ov c_{n,1} c_{n,2}=\ov c_{n,1}(2\pi) c_{n,2}(2\pi).
\ee
Let us write $z_n=z_{n,1}+iz_{n,2}, z_n^0=z_{n,1}^0+iz_{n,2}^0$,
where $z_{n,1},z_{n,2},z_{n,1}^0,z_{n,2}^0\in\R$, and
\be\la{zxy}
\quad \bz_n=\left(\!\!\!\ba{c}z_{n,1}\\z_{n,2}\ea\!\!\!\right),
\quad \bz_n^0=(z_{n,1}^0,z_{n,2}^0),
\qquad
 \fr{\pa}{\pa \bz_n^0}:=(\fr{\pa}{\pa z_{n,1}^0},\fr{\pa}{\pa z_{n,2}^0}), 
\qquad n\in\ov N.
\ee
Denote
\be\la{Zk}
\bZ=\left(\!\!\ba{c}\bz_1\\\vdots\\\bz_N\ea\!\!\right),\qquad \bZ^0=(\bz_1^0,\dots,\bz_N^0).
\ee
Then the differential $DP_*(X^0_*)$ 
of the Poincar\'e map $P_*$ on the factorspace $\cX/G$
is represented by the matrix
\be\la{Mder}
DP_*(X^0_*)=\left(
\ba{ccc}
\fr{\pa a}{\pa a^0}&\fr{\pa a}{\pa b^0}&\fr{\pa a}{\pa \bZ^0}\\\\
\fr{\pa b}{\pa a^0}&\fr{\pa b}{\pa b^0}&\fr{\pa b}{\pa \bZ^0}\\\\
\fr{\pa \bZ}{\pa a^0}&\fr{\pa \bZ}{\pa b^0}&\fr{\pa \bZ}{\pa \bZ^0}
\ea
\right).
\ee
Using  
 (\ref{1apr2rT}) with $\tau=2\pi$, we obtain that
 \beqn\la{Pmap2H}
 \!\!\!\!z_n\!\!&\!\!=\!\!&\!\!(\ov c_{n,1}^0
+i\ov c_{n,2}^0  2\pi \ov{\ti\om}_n^{(1)})
(-i c_{n,1}^0 2\pi \ov{\ti\om}_n^{(1)}
+c_{n,2}^0)+O(|\ti\om_n^{(1)}|^2)
\nonumber\\
\nonumber\\
\!\!\!\!&\!\!=\!\!&\!\!
z_n^0+
2\pi i\ov{\ti\om}_n^{(1)}[|c_{n,2}^0|^2-|c_{n,1}^0|^2]+O(|\ti\om_n^{(1)}|^2)
\nonumber\\
\nonumber\\
\!\!\!\!&\!\!=\!\!&\!\!
z_n^0+
2\pi i
(\beta_n\ov\nu+\fr{\ga_n}2)
[|c_{n,2}^0|^2-|c_{n,1}^0|^2]+O(|\ti\om_n^{(1)}|^2).
 \eeqn
We must calculate  the matrix (\ref{Mder}) at the 
ground states (\ref{zsol}) which correspond to the single point
\be\la{zsol2}
X^0_*=(0,0,0).
\ee
The formulas (\ref{Pmap4}) for the Poincar\'e map  become
\be\la{Pmap5}
 \left\{\ba{rcl}
 a&\approx&a_0(2\pi)+
 \sum_n\al_n
\rIm\Big\{z_n^0A_1\Big\}
\\\\
  &&+
   \sum_n\al_n\beta_n[|c_{n,2}^0|^2 -|c_{n,1}^0|^2]\rRe\big\{
 \ov \nu A_2\big\}
+
\fr12 \sum_n\al_n \ga_n
[|c_{n,2}^0|^2-|c_{n,1}^0|^2] 
A_3
 \\\\
b&\approx&\dot a_0(2\pi)+
 \sum_n\al_n
\rIm\Big\{z_n^0B_1\Big\}
\\\\
  &&+
   \sum_n\al_n\beta_n[|c_{n,2}^0|^2 -|c_{n,1}^0|^2]\rRe\big\{
 \ov \nu B_2\big\}
+
\fr12 \sum_n\al_n \ga_n
[|c_{n,2}^0|^2-|c_{n,1}^0|^2] 
B_3
 \ea\right|.
 \ee
Here $a_0(2\pi)$ and $\dot a_0(2\pi)$  are given by (\ref{MB31in}), so
using
(\ref{funs0}) 
and neglecting errors
$O(\vka^2)=O(10^{-14})$, we obtain
\be\la{ab0}
\left\{\ba{rcl}
a_0(2\pi)&=&a^0\dot E(2\pi)+[b^0+2\vka a^0]E(2\pi)
\approx a^0(1-2\pi\vka)
\\\\
\dot a_0(2\pi)&=&a^0\ddot E(2\pi)+[b^0+2\vka a^0]\dot E(2\pi)
\approx-2\vka a^0+[b^0+2\vka a^0](1-2\pi\vka)
\approx b^0(1-2\pi\vka)
\ea\right|.
\ee 
 At the point (\ref{zsol2}), we have by (\ref{c20}),
\be\la{zxy2}
\fr{\pa}{\pa \bz_l^0}\Big|_{z_l^0=0}
|c_{n,1}^0|^2=\fr{\pa}{\pa \bz_l^0}\Big|_{z_l^0=0}|c_{n,2}^0|^2=(0,0).
%
\ee
Differentiating the formulas (\ref{nu}) 
at the point (\ref{zsol2}), we obtain that
neglecting errors $O(\vka^2)=O(10^{-14})$,
we have
\be\la{nud}
 \fr{\pa \ov\nu}{\pa a^0}=-\fr i2+
\fr\vka4(1+2\pi i),
\qquad  \fr{\pa \ov\nu}{\pa b^0}=\fr12-\fr\vka4(2 \pi-i)=i\fr{\pa\ov \nu}{\pa a^0}.
\ee
Moreover, formulas (\ref{nu}) and (\ref{zxy2}) imply that 
at the point (\ref{zsol2}), we have
\be\la{nud2}
\fr{\pa \ov\nu}{\pa \bz_{n'}^0}=(0,0).
\ee
\smallskip

Now we can calculate the matrix (\ref{Mder}) at the point (\ref{zsol2}).
Differentiating the first  formula of  (\ref{Pmap5}),
and taking into account  (\ref{ab0})--(\ref{nud2}), 
we obtain that the first row of (\ref{Mder}) is 
\be\la{a-der}
\fr{\pa a}{\pa a^0}\!\approx \!1\!-\!2\pi\vka
\!-\!\bS[\fr {A_{22}}2\!+\!\fr\vka4(A_{21}\!-\!2\pi A_{22})],
\,\,
\fr{\pa a}{\pa b^0}\!\approx\! -\!\bS[\fr {A_{21}}2\!-\!\fr\vka4(2\pi A_{21}\!+\!A_{22})],
\,\,
 \fr{\pa a}{\pa \bz_{n'}^0}\!\approx\!\al_{n'}(A_{12},A_{11} ),
\ee
where $\bS$ is the summation (\ref{SSig}).
 Similarly, using (\ref{B123}), we obtain the second row 
 \be\la{b-der}
\fr{\pa b}{\pa a^0}\!\approx\!
 -\bS[\fr{B_{22}}2\!+\!\fr\vka4(B_{21}\!-\!2\pi B_{22})],\,\,
\fr{\pa b}{\pa b^0}\!\approx\! 1\!-\!2\pi\vka
\!-\!
 \bS[ \fr{B_{21}}2 \! -\!\fr\vka4(2\pi B_{21}\!+\!B_{22})] ,\,\,
 \fr{\pa b}{\pa \bz_{n'}^0}\!\approx\! \al_{n'}(B_{12},B_{11}),
\ee 
where $B_{11}=\rRe B_1$, $B_{12}=\rIm B_1$, and similarly for 
$B_{21}$ and $B_{22}$.
  Finally, 
  differentiating (\ref{Pmap2H})
 at the point (\ref{zsol2}), and using (\ref{zxy2}), (\ref{nud}), 
 we get
   \beqn\la{Z-der}
\fr{\pa z_n}{\pa a^0}&\approx&
2\pi i\beta_n
\fr{\pa\ov\nu}{\pa a^0}
=2\pi i\beta_n
[-\fr i2+
\fr\vka4(1+2\pi i)]=
 \pi\beta_n[1-\fr\vka2(2\pi-i)] 
 \nonumber\\
 \nonumber\\
 \fr{\pa z_n}{\pa b^0}&\approx&
 2\pi i\beta_n
\fr{\pa\ov\nu}{\pa b^0}
=2\pi i\beta_n
[\fr12-\fr\vka4(2 \pi-i)]=
 \pi\beta_n[i-\fr\vka2(2\pi i+1)]
\eeqn
Similarly, using (\ref{zxy2})  and (\ref{nud2}), we obtain
\beqn\la{Z-der2}
 \fr{\pa z_n}{\pa \bz_{n'}^0}\approx \de_{nn'}(1, i).
 \eeqn
 The expressions (\ref{Z-der}) and (\ref{Z-der2}) can be written as vectors and matrices:
 \be\la{Z-der3}
\fr{\pa \bz_n}{\pa a^0}\approx
 \pi\beta_n \left(\!\!\ba{c}1-\vka\pi\\ \fr\vka2\ea\!\!\right),
 \qquad\fr{\pa \bz_n}{\pa b^0}
 \approx
 \pi\beta_n\left(\!\!\ba{c}-\fr\vka2\\1-\vka\pi\ea\!\!\right),
 \qquad
 \fr{\pa \bz_n}{\pa \bz_{n'}^0}\approx \de_{nn'}
  \left(\!\!\ba{cc} 
  1
  &0\\0&1\ea\!\!\right).
 \ee

 \setcounter{equation}{0}
\section{Block-matrix approximation}
All calculations below essentially rely on the constants
 (\ref{abg}) which correspond to the ruby laser.
 Let us denote the matrix
 \be\la{MVW}
M\!=\!\left(\!\!
\ba{cc}
\fr{\pa a}{\pa a^0}&\fr{\pa a}{\pa b^0}
\\\\
\fr{\pa b}{\pa a^0}&\fr{\pa b}{\pa b^0}
\ea
\!\!\right).
\ee
 Recall that 
 for the typical dipole moment (\ref{Pval})
 we have  
 by (\ref{probs22}), (\ref{probs22cr}) and   (\ref{abg}), that
 \be\la{Sum}
 \bS=\sum_n\al_n\beta_n\sim 10^{-5},\qquad \ga_n\approx 2\times10^{-7},\qquad \vka\approx 10^{-7}.
 \ee
  Rewrite the formulas (\ref{a-der})--(\ref{b-der}) 
  substituting the expressions (\ref{nud}), and
     neglecting  all  terms containing 
 $\vka \bS\sim 10^{-12}$ and $\vka^2\sim 10^{-14}$:
 \be\la{a-derr}
\fr{\pa a}{\pa a^0}\approx
  1-2\pi\vka+\fr 12 A_{22} \bS,
\qquad
\fr{\pa a}{\pa b^0}\approx 
-  \fr12 A_{21}\bS,
\ee
  \be\la{b-derr}
\fr{\pa b}{\pa a^0}\approx
- \fr12 B_{22} \bS,
 \qquad
\fr{\pa b}{\pa b^0}\approx
1-2\pi\vka
-   \fr12  B_{21} \bS.
\ee
Hence,
\be\la{Amat}
M=M(\vka,\bS)
\approx  
\left(
\ba{ll}
1-2\pi\vka-\fr 12 A_{22}  \bS
&
-  \fr12 A_{21}\bS
\\\\
 - \fr12 B_{22}  \bS
&
1-2\pi\vka-   \fr12 B_{21} \bS
\ea
\right).
 \ee

 Further, denote the matrices
\be\la{DVW}
D_n=
  \left(\!\!\ba{cc}  1&0\\\\0&1\ea\right)
  ,\qquad
V_n=\al_n
\left(
\ba{cc}
A_{12} &A_{11}
\\\\
B_{12}&B_{11}
\ea
\right),\qquad
W_n=\pi\beta_n 
\left(
\ba{cc}
1-\vka\pi&-\fr\vka2
\\\\
\fr\vka2&1-\vka\pi
\ea
\right).
\ee
Using (\ref{B123}) and (\ref{A1}), we obtain that
\be\la{DVW2}
V_n\approx 
\pi\al_n
\left(
\ba{cc}
1-\vka\pi & \fr\vka2
\\\\
-\fr\vka2&1-\vka\pi 
\ea\right)
\ee
Now formulas (\ref{a-der}), (\ref{b-der}), (\ref{Amat}), and
(\ref{Z-der3}) give  the following block-matrix
approximation for  the differential (\ref{Mder}):
\be\la{Mder3}
DP_*(X^0_*)\approx\left(
\ba{cccccc}
M&V_1&V_2&V_3&\dots&V_N
\\
W_1&D_1&0&0&\dots&0\\
W_2  &0 &D_2&0&\dots&0\\
W_3    &0&0&D_3&\dots&0\\
\vdots    &\vdots&\vdots&\vdots&\vdots&\vdots\\
W_N&0&0&0&\dots&D_N
\ea
\right),
\ee
which  is not symmetric generally.
\br\la{rtr}
\rm
Since $\al_n\sim 10^{-23}$, the matix is very close to 
its  triangle approximation
\be\la{Mder3tr}
DP_*(X^0_*)\approx\left(
\ba{cccccc}
M&0&0&0&\dots&0
\\
W_1&D_1&0&0&\dots&0\\
W_2  &0 &D_2&0&\dots&0\\
W_3    &0&0&D_3&\dots&0\\
\vdots    &\vdots&\vdots&\vdots&\vdots&\vdots\\
W_N&0&0&0&\dots&D_N
\ea
\right).
\ee
The corresponding eigenvalues 
coincide with the ones of the matrices  $M$ and $D_n$. In particular,
for the  eigenvalues of $M$ we have $|\mu-1|\sim 10^{-5}$ by (\ref{Amat})
and (\ref{probs22}). 
\er

\section{Reduction to polynomial equation}
\la{sred}

The spectral problem for the multipliers reads
\be\la{eigDP}
DP_*(X^0_*) \bV=\mu \bV,\qquad \bV=\left(
\ba{c}
\bv_0\\ \bv_1\\
\dots\\
\bv_N
\ea
\right),
\ee
where all $\bv_n\in\R^2$  and $\bV\ne 0$.
We will reduce the spectral problem 
to a polynomial equation.
Namely,
substituting here the approximation (\ref{Mder3}), we obtain the 
equivalent 
system
\be\la{eigDP2}
(M-\mu)\bv_0+\sum_1^N V_n\bv_n=0,\qquad
W_n\bv_0+(D_n-\mu)\bv_n=0,\quad n\in\ov N.
\ee
Since we seek for eigenvalues with $|\mu|>1$, we can write 
\be\la{eigDP3}
\bv_n=(\mu-D_n)^{-1}W_n\bv_0
\approx
\pi\beta_n 
\left(\!\!
\ba{cc}
\fr{1-\vka\pi}{\mu-1}&\fr{-\vka/2}{\mu-1}
\\\\
\fr{\vka/2}{\mu-1+2\pi^2\ga_n^2}&\fr{1-\vka\pi}{\mu-1+2\pi^2\ga_n^2}
\ea
\!\!\right)
\bv_0,
\ee
where $\bv_0\ne 0$.
 Substituting into the first equation of (\ref{eigDP2}), we obtain that 
$
\cM(\mu)\bv_0=0,
$
where
\beqn\la{cMa1}
\cM(\mu)&\approx&
(M\!-\!\mu)\!+\!\pi^2\sum_1^N \left[\al_n\beta_n
\left(
\ba{cc}
1\!-\!\vka\pi \!\!&\!\! \fr\vka2
\\\\
\!-\!\fr\vka2\!\!&\!\!1\!-\!\vka\pi 
\ea\right)
\left(\!\!
\ba{cc}
\fr{1\!-\!\vka\pi}{\mu\!-\!1}&\fr{\!-\!\vka/2}{\mu\!-\!1}
\\\\
\fr{\vka/2}{\mu\!-\!1\!+\!2\pi^2\ga_n^2}&\fr{1\!-\!\vka\pi}{\mu\!-\!1\!+\!2\pi^2\ga_n^2}
\ea
\!\!\right)
\right]
\nonumber\\
\nonumber\\
&=&
(M\!-\!\mu)\!+\!\pi^2
\left(
\ba{cc}
1\!-\!\vka\pi \!\!&\!\! \fr\vka2
\\\\
\!-\!\fr\vka2\!\!&\!\!1\!-\!\vka\pi 
\ea\right)
\left(\!\!
\ba{cc}
\fr{1\!-\!\vka\pi}{\mu\!-\!1}\bS&\fr{\!-\!\vka/2}{\mu\!-\!1}\bS
\\\\
\fr\vka2\sum_1^N 
\fr{\al_n\beta_n}
{\mu\!-\!1\!+\!2\pi^2\ga_n^2}
&(1\!-\!\vka\pi)\sum_1^N 
\fr{\al_n\beta_n}{\mu\!-\!1\!+\!2\pi^2\ga_n^2}
\ea
\!\!\right).
\eeqn
Thus, the characteristic equation for the matrix $DP_*(X^0_*)$ is equivalent to
\be\la{eigDP5}
\det \cM(\mu)=0.
\ee
Let us consider the multipliers for which
$
2\pi^2\fr{\ga_n^2}{\mu-1}\ll1.
$
The existence of such 
multipliers is suggested by Remark \ref{rtr}.
For such multpliers,
the last two sums in (\ref{cMa1}) can be calculated as follows. 
First,
\beqn\la{Su}
&&\sum_1^N 
\fr{\al_n\beta_n}{\mu\!-\!1\!+\!2\pi^2\ga_n^2}\!=\!\fr1{\mu\!-\!1}\sum_1^N 
\fr{\al_n\beta_n}{1\!+\!2\pi^2\fr{\ga_n^2}{\mu\!-\!1}}\!\approx\!
\fr1{\mu\!-\!1}\sum_1^N 
\al_n\beta_n(1\!-\!2\pi^2\fr{\ga_n^2}{\mu\!-\!1})
\nonumber\\
\nonumber\\
&=&\fr\bS{\mu\!-\!1}\!-\!\fr{2\pi^2}{(\mu\!-\!1)^2}\sum_1^N 
\al_n\beta_n\ga_n^2\!=\!\fr\bS{\mu\!-\!1}\!-\!\fr{2\pi^2}{(\mu\!-\!1)^2}
\fr2{\Om_p\hbar^2c}
\sum_1^N 
(\bP_n\cdot\bX(x_n))^2(\bP_n\cdot \bba_p(x_n))^2.
\eeqn
Second, 
by Remark \ref{rH}, we can apply 
the {\it law of  large numbers}
for weakly dependent random values \ci{IL1971}, and obtain that
 $\bS\approx\ov N\cE \al_n\beta_n$.
\be\la{Su2}
\sum_1^N 
(\bP_n\!\cdot\!\bX(x_n))^2(\bP_n\!\cdot\! \bba_p(x_n))^2
\!=\!N\Si,\qquad \Si:=\cE(\bP_n\!\cdot\!\bX(x_n))^2(\bP_n\!\cdot \!\bba_p(x_n))^2.
\ee
The last expectation can be calculated 
using the fact that random values $\rm (P^1,P^2,P^3)=\bP_n$, 
$\rm(X^1,X^2,X^3)=\bX(x_n)$ and $\rm (a^1,a^2,a^3)=\bba_p(x_n)$ 
are independent, so
\be\la{Sig}
\Si\!=\!\cE\!\sum_{i,l=1}^3 \!\rm(P^iX^i)^2(P^lX^{l'})^2
\!=\!\cE\!\!\sum_{i,i',l,l'=1}^3\!\! \rm P^iP^{i'}P^lP^{l'} X^iX^{i'} a^la^{l'}
\!=\!\sum_{i,i',l,l'=1}^3 \!\cE\rm  [P^iP^{i'}P^lP^{l'}] \,\,\cE [X^iX^{i'}]\,\,
 \cE [a^la^{l'}]
\ee
By our assumption {\bf H4}, the random vector $\bba_p(x_n)$ is distributed uniformly on the sphere of a radius $\rm a_p$, so
\be\la{Ea}
 \cE\rm [a^la^{l'}]=\fr13\de^{ll'}a_p^2.
\ee
Further, {\it let us conjecture} that the distribution of the 
random vector $\bX(x_n)$ is close to the 
one of the eigenmodes 
(\ref{eigcubo})
of rectangular cuboid $V=[0,l_1]\times[0,l_2]\times[0,l_3]$. 
In this case, similarly to (\ref{probs2}),
\be\la{EX}
\cE{\rm [X^iX^{i'}]=\de^{ii'}\cE\rm |X^i|^2=
\fr{\de^{ii'}}N\sum_{n=1}^N |\bX^i(x_n)|^2 \approx
\fr{\de^{ii'}}{|V_a|}\int_{V_a}|X^i|^2(x)dx}= \fr{\de^{ii'}}{3|V|}|a^i|^2,
\ee
where $a^i=a_k^i$  are the amplitudes from  (\ref{eigcubo}).
Substituting (\ref{Ea}) and (\ref{EX}) into (\ref{Sig}), we obtain
that
\be\la{Sig2}
\Si
=\fr1{9|V|}{\rm |a_p|^2\sum_{i,l=1}^3 \cE  [|P^i|^2|P^l|^2]} |a^i|^2
\ee
\br
{\rm
According to Berry's conjecture, 
in the case of ergodic geodesic flow,
the distribution of 
values of Laplacian
eigenfunctions with sufficiently large eigenvalues
can be Gaussian \ci{B1977, LU2022}, thus different from the ``sinusoidal distribution" 
of eigenmodes (\ref{eigcubo}).

}
\er

Let us  consider the following two cases separately.
\smallskip\\
{\bf I. Polycrystalline  medium.}
Let us  calculate $\rm\cE[|P^i|^2|P^l|^2]$. 
By our assumption {\bf H1},
the random vector $\rm(P^1,P^2,P^3)$ is distributed uniformly on the sphere
of radius $\rm|P|$. Hence,
integrating in spherical coordinates,
we obtain 
\be\la{PPP}
\rm\cE|P^1|^2|P^2|^2=\fr1{15}|P|^4,\qquad \cE(P^1)^4=\fr1{5}|P|^4.
\ee
Hence, 
\be\la{Sigm}
\Si
=\fr1{9|V|}{\rm|a_p|^2|P|^4\sum_{i=1}^3  (\fr15+\fr2{15})}|a^i|^2=
\fr1{27|V|}\rm|a_p|^2|\bP|^4
\ee
since $a=(a^1,a^2,a^3)$ is the unit vector.
\smallskip\\
{\bf II. Crystalline medium.}  
In this case
 the vectors $\bP_n=\bP$  do not depend on $n$.
Thus, (\ref{Sig2}) becomes
\be\la{Sig3}
\Si=
\fr1{9|V|}{\rm |a_p|^2\sum_{i,l=1}^3  [|P^i|^2|P^l|^2]} |a^i|^2
=\fr{{\rm a_p^2}|\bP|^2}{9|V|}\sum_{i=1}^3 |\bP^i|^2|a^i|^2.
\ee

In both cases,
   (\ref{eigDP5}) for $\mu\ne 1$
   is equivalent to the algebraic equation
\be\la{pol6}
p(\mu)=0,\qquad p(\mu)=\det[(\mu-1)^2\cM(\mu)],
\ee
where $p(\mu)$ is a polynomial of degree six.

\br\la{rsym}
{\rm
i) The summation $\bS$ reappears in the calculations (\ref{cMa1}) and (\ref{Su}).
\\
ii) The degree six of the polynomial can be reduced to three 
using the approximate symmetry of type (\ref{LP222}), see Remark \ref{rred}.
}
\er

\br\la{rM}
{\rm
For every nonzero eigenvector (\ref{eigDP}), its component $\bv_0$, 
representing the Maxwell field,
 does not vanish.
Hence, 
if (\ref{ph}) holds, then
the magnitude of the Maxwell field
provisionally 
 increases exponentially for large (but bounded) times.
 
}
\er

\br\la{rP6}
{\rm
 We expect that 
 \\
i) all the multipliers $|\mu|$ are simple with probability one;
\\
ii)  the multipliers form {\it clusters}  around the  roots of the polynomial $p(\mu)$;
\\
iii) the condition (\ref{ph}) holds for the 
roots of the polynomial $p(\mu)$ when 
\be\la{pumdam}
{\rm a}_p\ge d(\si_1),
\ee
where $d(\si_1)$ is the corresponding threshold, which is an increasing function of $\si_1$.
Physically, the condition (\ref{pumdam}) means that
the amplitude of the pumping exceeds the damping losses.
}
\er

\appendix

\setcounter{section}{0}
\setcounter{equation}{0}
\protect\renewcommand{\thetheorem}{\Alph{section}.\arabic{theorem}}

\setcounter{equation}{0}
\section{On averaging of  slow rotations}
Here we justify the 
approximation of the Schr\"odinger equations from (\ref{MB31c})
by their averaged version (\ref{avB}) (``rotating wave approximation''
\ci{AE1975,AT1955,H1984,N1973,R1937,SSL1978,SZ1997,S2010}). 
Let us show that 
this approximation is very accurate for small matrices 
$\Om_n(\tau)$. Namely, consider the system
 \be\la{B}
\dot c(\tau)=-i\Om(\tau) 
  c(\tau),\qquad \tau\in[0,T],
\ee
where  $\Om(\tau)$ is a 
bounded measurable  $2\times 2$ complex matrix-function,
\be\la{Mnor}
\ve:=\sup_{\tau\in[0,T]} | \Om(\tau)|<\infty.
\ee
Denote the averaged matrix
\be\la{hatom}
\ti\Om:=\fr1T\int_0^T\Om(\tau)d\tau
\ee
and consider the corresponding averaged system 
 \be\la{Bav}
\dot {\hat c}(\tau)=-i\ti\Om 
  \hat c(\tau),\qquad \tau\in[0,T].
\ee

\bl\la{lA} 
For solutions of (\ref{B}) and (\ref{Bav}) with identical initial data
$\hat c(0)=c(0)=c_0$, 
their final values $c(T)$ and 
$\hat c(T)$ are very close for small $\ve$:
\be\la{ccc} 
|c(T)-\hat c(T)|=O(\ve^2),\qquad \ve\to 0.
\ee
\el
\bpr
Rewrite (\ref{B}) as
\be\la{Bs}
\dot c_\ve(\tau)=-i\ve\Om_1(\tau) 
  c_\ve(\tau),\qquad \tau\in[0,T];\qquad c_\ve(0)=c_0,
\ee
where $\Om_1(\tau):=\fr1\ve\Om(\tau)$.
The corresponding averaged system (\ref{Bav}) now reads
\be\la{Bavs}
\dot {\hat c}_\ve(\tau)=-i\ve\hat \Om_1(\tau) 
 \hat c_\ve(\tau),\qquad \tau\in[0,T];\qquad \hat c_\ve(0)=c_0,
\ee
where $\ti\Om_1(\tau):=\ds\fr1T\int_0^T\Om_1(\tau)d\tau$.
It suffices to prove that
\be\la{ccc2} 
|c_\ve(T)-\hat c_\ve(T)|=O(\ve^2),\qquad \ve\to 0.
\ee
The Taylor formula gives
\be\la{Texp}
c_\ve(\tau)=c_0+\ve[\pa_\ve c_\ve(\tau)]|_{\ve=0}+O(\ve^2),
\qquad
\hat c_\ve(\tau)=c_0+\ve[\pa_\ve \hat c_\ve(\tau)]|_{\ve=0}+O(\ve^2).
\ee
Therefore,
(\ref{ccc2}) will follow from the identity
\be\la{devep}
[\pa_\ve c_\ve(T)]|_{\ve=0}=[\pa_\ve \hat c_\ve(T)]|_{\ve=0}.
\ee
Indeed, $\hat c_\ve(T)=e^{-i\ve\ti\Om T} c_0$, so
the right-hand side of (\ref{devep}) is equal to
\be\la{rhs}
[\pa_\ve \hat c_\ve(T)]|_{\ve=0}=-i\ti\Om_1 T c_0=-i\int_0^T \Om_1(\tau)c_0d\tau.
\ee
It remains to calculate the left-hand side. 
Denote 
\be\la{bve}
b_\ve(\tau):=\pa_\ve c_\ve(\tau).
\ee
Differentiating equation (\ref{Bs}) in $\ve$,
we obtain 
\be\la{Bsd}
\dot b_\ve(\tau)=-i\Om_1(\tau) 
  c_\ve(\tau)-i\ve\Om_1(\tau) 
  b_\ve(\tau),\qquad \tau\in[0,T];\qquad b_\ve(0)=0.
\ee
In particular, for $\ve=0$ we have 
\be\la{Bsd2}
\dot b_0(\tau)=-i\Om_1(\tau) 
  c_0,\qquad \tau\in[0,T];\qquad b_0(0)=0.
\ee
Hence,
\be\la{Bsd3}
b_0(T)=-i\int_0^T\Om_1(\tau) 
  c_0 d\tau.
\ee
Now (\ref{devep}) is proved.
\epr

\br
{\rm
Equation (\ref{Bs})  
 can be rewritten for $\ti c_\ve(s):=c_\ve(\fr s\ve)$
as
\be\la{Bsr}
\pa_s \ti c_\ve(s)=-i\Om_1(\fr s\ve) 
  \ti c_\ve(s),\qquad s\in[0,\ve T],
\ee
where $s=\ve\tau$ is the `slow time'.
Thus, Lemma \ref{lA} is a  specific version of  averaging principle.
}
\er

\setcounter{equation}{0}
\section{Some integrals}\la{aB}

\subsection{Integrals (\ref{I12})}

First, (\ref{funs0}) implies   that 
\beqn\la{I1}
I_1(\tau)&:=&
 \int_0^\tau e^{-i\tau'}
 e^{-\vka(\tau-\tau')}\sin(\tau-\tau') d\tau'
=\int_0^\tau e^{-i\tau'}
 [e^{(-\vka+i)(\tau-\tau')}-e^{(-\vka-i)(\tau-\tau')}]
d\tau'
\nonumber\\
\nonumber\\
&=&\fr1{2i}e^{(-\vka+i)\tau}
\int_0^\tau 
 e^{(\vka-2i)\tau'}d\tau'-
 \fr1{2i}e^{(-\vka-i)\tau}
\int_0^\tau 
 e^{\vka\tau'}
d\tau'
\nonumber\\
\nonumber\\
&=&\fr1{2i}e^{(-\vka+i)\tau}
\fr
{e^{(\vka-2i)\tau}-1}{\vka-2i}-
 \fr1{2i}e^{(-\vka-i)\tau}
\fr
{e^{\vka\tau}-1}{\vka}
\nonumber\\
\nonumber\\
&=&
\fr
{e^{-i\tau}-e^{(-\vka+i)\tau}}{2i\vka+4}-
\fr
{e^{-i\tau}-e^{(-\vka-i)\tau}}{2i\vka}
\nonumber\\
\nonumber\\
&\approx&
-\fr i2\sin\tau-
\fr\tau{2i}e^{-i\tau}+\vka\Big[\fr{-\sin\tau+\tau e^{i\tau}}{4}+\fr{\tau^2e^{-i\tau}}{4i}\Big],
\eeqn
where we have neglected errors $O(\vka^2)=O(10^{-14})$. 
Similarly,
\beqn\la{I2}
I_2(\tau)&:=&
\fr{1}2
 \int_0^\tau \tau'
\cos\tau'
 e^{-\vka(\tau-\tau')}\sin(\tau-\tau')d\tau'
  =\fr{1}4\int_0^\tau \tau'
  e^{-\vka(\tau-\tau')}\Big[
  \sin\tau+\sin(\tau-2\tau')
  \Big]
    d\tau'
    \nonumber\\
  \nonumber\\
  &=&
  \fr{1}4 e^{-\vka\tau}
  \Big[\sin\tau\int_0^\tau \tau'e^{\vka\tau'} d\tau'+
  \rIm\int_0^\tau \tau'e^{\vka\tau'}e^{i(\tau-2\tau')} d\tau'\Big]
  \nonumber\\
  \nonumber\\
  &=&
  \fr{1}4 e^{-\vka\tau}
  \Big[\sin\tau\Big(\tau \fr{e^{\vka\tau}}{\vka}-\int_0^\tau \fr{e^{\vka\tau'}}{\vka}d\tau' \Big)
  +
  \rIm\Big(e^{i\tau}
  \tau \fr{e^{(\vka-2i)\tau}}{\vka-2i}
  -e^{i\tau}
  \int_0^\tau \fr{e^{(\vka-2i)\tau'}}{\vka-2i}d\tau' \Big)\Big]
   \nonumber\\
  \nonumber\\
  &=&
  \fr{1}4 e^{-\vka\tau}
  \Big[\sin\tau\Big(\tau \fr{e^{\vka\tau}}{\vka}-
  \fr{e^{\vka\tau}-1}{\vka^2} \Big)
  +
  \rIm\Big(e^{i\tau}
  \tau\fr{e^{(\vka-2i)\tau}}{\vka-2i}
  -e^{i\tau}\fr
{  e^{(\vka-2i)\tau}-1}{(\vka-2i)^2}\Big)\Big]
   \nonumber\\
  \nonumber\\
  &=&
  \fr{1}4 e^{-\vka\tau}
  \Big[\sin\tau(\fr{\vka\tau e^{\vka\tau}-
  e^{\vka\tau}+1 )
}{\vka^2} 
+
  \rIm\Big(
  \tau\fr{e^{(\vka-i)\tau}(\vka+2i)}{\vka^2+4}
  -\fr
{  (e^{(\vka-i)\tau}-e^{i\tau})(\vka+2i)^2}{(\vka^2+4)^2}\Big)\Big]
  \nonumber\\
  \nonumber\\
  &\approx&
  \fr{1}4 e^{-\vka\tau}
  \Big[\sin\tau
  (\fr{\tau^2}2+\fr{\vka\tau^3}2)+
  \tau \fr{e^{\vka\tau}}4
  \rIm(e^{-i\tau}(\vka+2i))
  -\fr1{16}\rIm
  [(e^{(\vka-i)\tau}-e^{i\tau})(\vka+2i)^2]\Big]
  \nonumber\\
  \nonumber\\
  &\approx&
  \fr{1}4 e^{-\vka\tau}
  \Big[\sin\tau
  (\fr{\tau^2}2+\fr{\vka\tau^3}2)+
  \tau \fr{e^{\vka\tau}}4
  [2\cos\tau-\vka\sin\tau]
  \nonumber\\
  \nonumber\\
  &&
    -\fr1{16}[
    (e^{\vka\tau}\cos\tau-\cos\tau)4\vka
    -4(-e^{\vka\tau}\sin\tau-\sin\tau)
  ] \Big]
    \nonumber\\
  \nonumber\\
  &\approx&
  \fr{1}4 e^{-\vka\tau}
  \Big[\sin\tau
  (\fr{\tau^2}2+\fr{\vka\tau^3}2)+
  \fr\tau4 [e^{\vka\tau}
  2\cos\tau-\vka\sin\tau]
    -\fr1{4}
    (e^{\vka\tau}+1)\sin\tau
  \Big]
    \nonumber\\
  \nonumber\\
  &\approx&
  \fr{1}4 
  \Big[\sin\tau
  \fr{\tau^2}2+
  \fr\tau4 [
  2\cos\tau-\vka\sin\tau]
    -\fr1{4}
    (1+e^{-\vka\tau})\sin\tau
  \Big]
   \nonumber\\
  \nonumber\\
  &\approx&
  \fr{1}4 
  \Big[\sin\tau
  \fr{\tau^2}2+
  \fr\tau4 [
  2\cos\tau-\vka\sin\tau]
    -\fr1{4}
    (2-\vka\tau)\sin\tau
  \Big]
   \nonumber\\
  \nonumber\\
  &=&
  \fr{1}8
  \Big[\tau^2\sin\tau
  +
  \tau
  \cos\tau
    -
    \sin\tau
  \Big].\qquad 
  \eeqn

\subsection{Integrals (\ref{JI12})}

Calculate integrals 
\beqn
\int_0^{2\pi} \tau e^{i\tau}d\tau&=&-2\pi i\approx -6.28i,\qquad
\int_0^{2\pi} \tau\cos\tau d\tau=0,\qquad
\int_0^{2\pi} \tau e^{2i\tau}d\tau=-\pi i,
\nonumber\\
\nonumber\\\nonumber
\int_0^{2\pi} \tau^2 e^{i\tau}d\tau&=&-4\pi^2i+i\int_0^{2\pi} 2\tau e^{i\tau}d\tau=-4\pi^2i +4\pi,
\nonumber\\
\nonumber\\\nonumber
\int_0^{2\pi} \tau^2 e^{2i\tau}d\tau&=&-2\pi^2i+i\int_0^{2\pi} \tau e^{2i\tau}d\tau=-2\pi^2i +\pi,
\nonumber\\
\nonumber\\\nonumber
\int_0^{2\pi} \tau^2\sin\tau d\tau&=&-4\pi^2i.
\eeqn
Then, neglecting errors $O(\vka^2)=O(10^{-14})$, we can
rewrite (\ref{JI12}) as
\be\la{JI12c}
\left\{\!\!\!\!\ba{rcl}
J_1&:=&\ds\fr1{2\pi}\int_0^{2\pi}  I_1'(\tau)e^{-i\tau}d\tau
\approx \ds\fr1{2\pi}\int_0^{2\pi} 
\Big[
-\fr i2\sin\tau-
\fr\tau{2i}e^{-i\tau}+\vka\big[\fr{-\sin\tau+\tau e^{i\tau}}{4}+\fr{\tau^2e^{-i\tau}}{4i}\big]
\Big]'e^{-i\tau}d\tau
\\\\
&=&\ds\fr1{2\pi}\int_0^{2\pi} 
\Big[
-\fr i2\cos\tau-
\fr{ e^{-i\tau}-i\tau e^{-i\tau}}{2i}+\vka\big[\fr{-\cos\tau+ e^{i\tau}+i\tau e^{i\tau}}{4}+\fr{2\tau e^{-i\tau}-i\tau^2e^{-i\tau}}{4i}\big]
\Big]e^{-i\tau}d\tau
\\\\
&=&\ds\fr1{2\pi}\int_0^{2\pi} 
\Big[
-\fr i2\cos\tau e^{-i\tau}-
\fr{ e^{-i2\tau}-i\tau e^{-2i\tau}}{2i}+\vka\big[\fr{-\cos\tau e^{-i\tau}+i\tau}{4}+\fr{2\tau e^{-2i\tau}-i\tau^2e^{-2i\tau}}{4i}\big]
\Big]d\tau
\\\\
&=&\ds\fr1{2\pi}\int_0^{2\pi} 
\Big[
-\fr i4+
\fr{\tau e^{-2i\tau}}{2}+\vka\big[\fr{-\fr12+i\tau}{4}+\fr{2\tau e^{-2i\tau}-i\tau^2e^{-2i\tau}}{4i}\big]
\Big]d\tau
\\\\
&=& 
\ds
-\fr i4+
\fr{\pi i}{4\pi}+\vka\big[\fr{-\fr12+i\pi}{4}+\fr{2\pi i-i(2\pi^2i +\pi)}{8\pi i}\big]
%
= 0,
\\\\
J_2\!\!&\!\!:=\!\!&\!\!
\ds\fr1{2\pi}\int_0^{2\pi}  I_2'(\tau)e^{-i\tau}d\tau
\approx
\ds\fr1{16\pi}\int_0^{2\pi} 
\Big[\tau^2\sin\tau
  +
  \tau
  \cos\tau
    -
    \sin\tau
  \Big]'e^{-i\tau}d\tau
  \\\\
  &=&\ds\fr1{16\pi}\int_0^{2\pi} 
\Big[\tau\sin\tau+\tau^2\cos\tau
  \Big]e^{-i\tau}d\tau =\ds\fr1{32\pi}\int_0^{2\pi} 
\Big[-i\tau(1-e^{-2i\tau})+\tau^2(1+e^{-2i\tau})
  \Big]d\tau 
  \\\\
   &=&\ds\fr1{32\pi}
\Big[-i2\pi^2-\pi+\fr {8\pi^3}3+2\pi^2i +\pi
  \Big]=\fr{\pi^2}{12}.
  \ea\right|.
\ee

\subsection{Integrals (\ref{AB})}
Neglecting errors $O(\vka^2)=O(10^{-14})$, we obtain
 \beqn\la{A1}
 A_1&\approx&
  -\int_0^{2\pi} e^{-i\tau'} 
  e^{-\vka(2\pi-\tau')}\sin\tau'd\tau' 
  =-\fr{e^{-2\pi\vka}}{2i}
  \int_0^{2\pi} e^{-i\tau'} 
  e^{\vka\tau'}[e^{i\tau'}-e^{-i\tau'}]d\tau'
  \nonumber\\
  \nonumber\\
  &=&
  -\fr{e^{-2\pi\vka}}{2i}
  \int_0^{2\pi} 
  [e^{\vka\tau'}-e^{(\vka-2i)\tau'}]d\tau'
 =-\fr{e^{-2\pi\vka}}{2i}
  \Big[
  \fr{e^{2\pi\vka}-1}{\vka}-
   \fr{e^{2\pi\vka}-1}{\vka-2i}\Big]
   \nonumber\\
  \nonumber\\
&  \approx&
  -\fr{e^{-2\pi\vka}}{2i}
  \Big[
  \fr{2\pi\vka+\fr12(2\pi\vka)^2}{\vka}-
   \fr{2\pi\vka}{4}(\vka+2i)\Big]
   \approx-\fr{e^{-2\pi\vka}}{2i}
  \Big[
  2\pi+2\vka\pi^2-
   \vka\pi i\Big]
     \nonumber\\
  \nonumber\\
&  \approx&
  (1-2\pi\vka)
  \Big[
  \pi i+\vka\pi^2i+
   \fr{\vka\pi}{2}\Big]
  \approx
  \pi i+\vka\pi^2i+
   \fr{\vka\pi}{2}-\pi^22\vka i
   \nonumber\\
  \nonumber\\
  &=&
  \pi i+\vka[-\pi^2i+
   \fr{\pi}{2}]
   =
  A_{11}+iA_{12},\qquad A_{11}=\vka   \fr{\pi}{2},\quad
  A_{12}=\pi- \vka\pi^2. 
 \eeqn
 Similarly,
 \beqn\la{A2}
~ \!\!\!\!\!\!\!\!&\!\!\!\!\!\!\!\!&\!\!\!\!A_2\!\approx\!
 -\fr{e^{-\pi\si_1}}{2i}
  \int_0^{2\pi} \tau'
  [e^{\vka\tau'}-e^{(\vka-2i)\tau'}]d\tau'
  \nonumber\\
  \nonumber\\
 \!\!\!\!&\!\!\!\! =\!\!\!\!&\!\!\!\!
  -\fr{e^{\!-\!\pi\si_1}}{2i}
  \Big[\tau
  \fr{e^{\vka\tau}}{\vka}\!-\!
   \fr{e^{\vka\tau}}{\vka^2}
 \! -\!
  \tau
  \fr{e^{\vka\tau}}{\vka\!-\!2i}\!+\!
   \fr{e^{\vka\tau}}{(\vka\!-\!2i)^2}\Big]\Big|_0^{2\pi}
  \!=\!
-\fr{e^{-\pi\si_1}}{2i}
  \Big[\fr{\tau\vka\!-\!1}{\vka^2}  e^{\vka\tau}
  \!-\!
  \fr{\tau (\vka\!-\!2i)\!-1}{(\vka\!-\!2i)^2} e^{\vka\tau}
   \Big]\Big|_0^{2\pi}
    \nonumber\\
  \nonumber\\
 ~\!\!\!\! &\!\!\!\!\approx\!\!\!\!&\!\!\!\!
    -\fr{e^{\!-\!\pi\si_1}}{2i}
  \Big[\fr{\tau\vka\!-\!1}{\vka^2}  e^{\vka\tau}
  \!-\!
  \fr{\tau (\vka\!-\!2i)\!-\!1}{16} (\vka\!+\!2i)^2e^{\vka\tau}
   \Big]\Big|_0^{2\pi}
      \nonumber\\
  \nonumber\\
  ~\!\!\!\!&\!\!\!\!=\!\!\!\!&\!\!\!\!
    -\fr{e^{-\pi\si_1}}{2i}
  \Big[\fr{(\pi\si_1-1) e^{\pi\si_1}+1}{\vka^2} 
  -
  \fr{(\vka+2i)^2}{16}
  [( \pi\si_1-4\pi i-1)e^{\pi\si_1}+1]
   \Big]
   \nonumber\\
  \nonumber\\
  ~\!\!\!\!&\!\!\!\!\approx\!\!\!\!&\!\!\!\! -\fr{e^{-\pi\si_1}}{2i}
  \Big[\fr{(\pi\si_1\!-\!1) 
  (1\!+\!\pi\si_1\!+\!\fr12 (\pi\si_1)^2\!+\!\fr16 (\pi\si_1)^3)
   \! +\!1}{(\vka)^2} 
 \! -\!\pi i \! -\!\fr{\pi\si_1}{2}  \!+ \! \fr{1}{4}[ \pi\si_1\!+\!(\!-\!4\pi i\!-\!1){\pi\si_1}]
   \Big]
    \nonumber\\
  \nonumber\\
  \!\!\!\!&\!\!\!\!\approx\!\!\!\!&\!\!\!\!
    -\fr{e^{\!-\!\pi\si_1}}{2i}
  \Big[\fr{
  (\pi\si_1)^2+
  (\pi\si_1\!-\!1) 
  (\fr12 (\pi\si_1)^2\!+\!\fr16 (\pi\si_1)^3)}{\vka^2} 
  \!-\!\pi i  \!-\!\fr{\pi\si_1}{2} \! +  \!\fr{1}{4}[ \pi\si_1\!+\!(\!-\!4\pi i\!-\!1){\pi\si_1}]
   \Big]
    \nonumber\\
  \nonumber\\
 \!\!\!\! &\!\!\!\!=\!\!\!\!&\!\!\!\!
    -\fr{e^{-\pi\si_1}}{2i}
  \Big[{(4\pi^2+(\pi\si_1-1) 
  (  2 \pi^2+\fr23 \pi^3\si_1)}
   -\pi i  -\fr{\pi\si_1}{2}    -\pi^2 \si_1 i]
   \Big]
    \nonumber\\
  \nonumber\\
  \!\!\!\!&\!\!\!\!\approx\!\!\!\!&\!\!\!\!
    -\fr{e^{\!-\!\pi\si_1}}{2i}
  \Big[2 \pi^2\!-\!\pi i
 \! +\!\si_1[\fr43\pi^3\!-\!\fr\pi2\!-\!\pi^2 i]
   \Big]
   \! =\!
    {e^{\!-\!\pi\si_1}}
  \Big[\pi^2i\!+\!\fr\pi 2
  \!+\!\si_1[\fr23\pi^3i\!-\!\fr\pi4 i\!+\!\fr{\pi^2} 2]
   \Big]
   \nonumber\\
  \nonumber\\
  &=&
  \Big[\pi^2i+\fr\pi 2
  +\si_1[\fr23\pi^3i-\fr\pi4 i+\fr{\pi^2} 2]
   \Big]-\pi\si_1\Big[ \pi^2i+\fr\pi 2
   \Big]
   \nonumber\\
  \nonumber\\
  &=\!\!\!\!&\!\!\!\!
  \pi^2i\!+\!\fr\pi 2
  \!-\!\si_1 i[\fr{\pi^3}3\!+\!\fr\pi4]
%
  \!=\!
  A_{21}\!+\!iA_{22},\,\,\,A_{21}\!=\!\fr\pi 2,\,\,
  A_{22}\!=   \pi^2 \! -\!\si_1 [\fr{\pi^3}3\!+\!\fr\pi4].
 \eeqn
 Hence, 
 \be\la{A3}
  A_3=\rRe A_2\approx    \pi^2
  -\si_1 [\fr{\pi^3}3+\fr\pi4].
 \ee
Finally,
\be\la{B123}
\left\{\ba{rcl}
 B_1&=&
 - \ds i\int_0^{2\pi} e^{-i\tau'}  E(2\pi-\tau')d\tau'=-iA_1
 \\\\
 B_2
 &=&\ds\int_0^{2\pi} [ e^{-i\tau'}  -i\tau' e^{-i\tau'} ]  E(2\pi-\tau')d\tau'=
 A_1-iA_2
 \\\\
 B_3
 &=&\rRe B_2=
A_{11}+A_{22}
\ea\right|.
  \ee


\end{document}